\documentclass[twocolumn,showpacs,preprintnumbers,prd,nofootinbib]{revtex4-1}
\usepackage{feynman}
\usepackage{slashed}
\usepackage{graphicx,amsmath,amssymb,xcolor,hyperref}

\def\MW{M_{\rm EW}}

\begin{document}

\author{Andreas Bally}
\email{andreas.bally@mpi-hd.mpg.de}
\author{Yi Chung}
\email{yi.chung@mpi-hd.mpg.de}
\author{Florian Goertz}
\email{florian.goertz@mpi-hd.mpg.de}
\affiliation{Max-Planck-Institut f\"ur Kernphysik, Saupfercheckweg 1, 69117 Heidelberg, Germany}

\title{The Hierarchy Problem and the Top Yukawa:\\
An Alternative to Top Partner Solutions}

\begin{abstract}

We discuss the role of the top-quark Yukawa coupling $y_t$ concerning the hierarchy problem and construct an alternative scheme to the conventional solutions with top partners. In traditional models, like SUSY or composite Higgs, top partners cancel the top loop contribution to the Higgs quadratic term. The lack of evidence for such colored partners however drives these models into more and more fine-tuned regions. Here, an alternative means to mitigate the top loop, allowing for natural electroweak symmetry breaking, is presented. Emphasizing that we have not measured the top-Higgs interactions at high scales yet, we envisage scenarios where this interaction is only approaching its sizable strength in the infra-red, but gets strongly suppressed at high scales. We first discuss possible effects via a modification of the running of the top Yukawa coupling. Then, we turn to models where the top Yukawa is generated at one-loop level. Originated from a dimension-six operator, it drops when crossing the mass threshold of new degrees of freedom. In either case, the top partners are replaced by some new top-philic particles with strong interaction. Thus, a very different phenomenology, such as large top mass running and signals in four top final states, is introduced, which will be discussed in detail. With the assistance of this mechanism, the solution to the hierarchy problem can be pushed to a (well-defined) higher scale, and a final test of naturalness might be deferred to a 100 TeV Collider, like the FCC.

\end{abstract}

\maketitle

\section{Introduction}
The standard model (SM) of particle physics, once extended by new physics (NP) at higher scales, 
suffers generically from the hierarchy problem (HP), {\it i.e.}, the fact that the electroweak (EW) scale
$\MW \sim 100$\,GeV is radiatively unstable with respect to large mass scales.
In fact, without an additional peculiar structure, one would rather expect the mass of the 
Higgs boson (and thus the EW scale) to reside close to the highest scales in the theory, such 
as the scale of grand unification \mbox{$M_{\rm GUT} \sim 10^{15}$\,GeV} $\gg \MW$ or the Planck scale 
$M_{\rm Pl} \sim 10^{19}$\,GeV. The reason is that in the SM the Higgs mass squared 
parameter $m_H^2$ is not protected by a symmetry and thus is sensitive to corrections from any heavy 
particle with mass $M \gg \MW$, coupled to the Higgs field. This drives $m_H^2 \to \lambda_{\rm NP} M^2$, with 
$\lambda_{\rm NP}$ a product of couplings (including loop factors $\sim (4 \pi)^{-2}$), 
describing the interaction strength of the NP with the Higgs sector.

Considering the SM as an effective theory, augmented with dimension $D>4$ operators 
${\cal O}_D^{(i)}/\Lambda_{\rm NP}^{D-4}$, one would indeed generically expect the coefficient of the 
$D=2$ Higgs-squared operator $|H|^2$ to reside close to 
the cutoff squared of the low-energy theory $\Lambda_{\rm NP}^2$, defined by the scale where new particles enter
\begin{equation}
\label{eq:cor}
m_H^2 \sim \lambda_{\rm NP} \Lambda_{\rm NP}^2\,,
\end{equation}
with $\Lambda_{\rm NP}\sim M$.
Above the scale $\Lambda_{\rm NP}$, on the other hand, the new physics could potentially cure the HP, in a sense that
the corrections to the Higgs mass are cutoff at this scale, {\it e.g.}, via supersymmetry (SUSY) or a composite pseudo-Goldstone Higgs \cite{Kaplan:1983fs, Kaplan:1983sm,Dugan:1984hq}. 
Still, in case that $\lambda_{\rm NP} \Lambda_{\rm NP}^2 \gg \MW^2$ at least a {\it little} HP remains with a residual fine-tuning between the bare Higgs mass and quantum corrections of $\Delta_{\rm FT} \sim \MW^2/(\lambda_{\rm NP} \Lambda_{\rm NP}^2)$, required to keep the Higgs boson light. 

Because at the LHC we are just surpassing the weak scale and so far no NP has been found, experiment renders the HP more acute than ever.
We note that, even if the NP does not couple directly to the Higgs sector, 
given that it couples in some way to the SM, $\lambda_{\rm NP}$ can not be arbitrarily small. Since natural solutions to the HP, as mentioned before, often invoke some new symmetry making the Higgs-mass corrections cancel in the symmetric limit, they generically feature new particles related to SM particles by this symmetry, such as top partners, which cancel out the divergent loop corrections. \ \

In fact, regarding the known interactions, the large top Yukawa contribution (coming with an additional color factor) is most severe, such that the top partner is expected to be the lightest degree of freedom to maintain naturalness. Termed in symmetries, the top quark Yukawa coupling is the major contribution to the breaking of the Higgs shift symmetry, 
\begin{equation}
\label{eq:ssm}
H \to H + a\,,
\end{equation}
which - in the absence of large breaking - could be used to justify a small $m_H$.\footnote{The heavy top quark plays in fact the most prominent role in many frameworks beyond the SM, where it is often also strongly coupled to the NP sector.} 

However, after years of searches, the bounds on the mass of colored top partners have reached around $1500$ GeV for both scalar partners \cite{CMS:2020pyk, ATLAS:2020aci} and fermionic partners \cite{CMS:2019eqb, ATLAS:2018ziw,ATLAS:2022hnn,CMS:2022fck,ATLAS:2022ozf}. The absence of the top partners starts challenging the naturalness of this type of models, due to the required fine-tuning. 

This can be shown numerically by evaluating the one-loop Higgs-mass correction due to the top quark
\begin{align}\label{toploop0}
\Delta m_H^2|_{\text{top}}
&\sim-i\,2N_c\,y_t^2 \int_{}^{} \frac{d^4k}{(2\pi)^4}\frac{k^2+m_t^2}{(k^2-m_t^2)^2}\nonumber\\
&= -\frac{3}{8\pi^2}y_t^2
\left[\Lambda_{\rm NP}^2-3\,m_t^2\,\text{ln}\left(\frac{\Lambda^2_{\rm NP}}{m_t^2}\right)+\cdots\right]~,
\end{align}
where we only kept the $\Lambda_{\rm NP}$-dependent terms.\footnote{We identify the cutoff with the NP scale, which would lead to corresponding finite corrections in the presence of new thresholds. These can be accompanied by another loop factor if the NP couples indirectly via the top-loop. This might be (partially) lifted in case of the NP coupling more strongly or interacting directly with the Higgs via the top Yukawa, the latter potentially fixed by (super)symmetry.} The presence of top partners, introducing $\Delta m_H^2|_{\text{top partner}}$, restores the symmetry and guarantees the cancellation of the quadratically UV sensitive term $\sim \Lambda_{\rm NP}^2$:
\begin{align}\label{toploop+}
\Delta m_H^2|_{\text{top}}+\Delta m_H^2|_{\text{top partner}}
\sim -\frac{3}{8\pi^2}y_t^2 M_T^2\,\text{ln}\left(\frac{\Lambda_{\rm NP}^2}{M_T^2}\right)~,
\end{align}
where $M_T$ is the mass of the top partner. The logarithm is negligible for generic composite Higgs models (CHM) and low-scale SUSY models. 
The observed Higgs mass and vacuum expectation value (VEV) result in $m_H^2=-\left(88\text{ GeV}\right)^2$. The generic estimation of the correction should be of the same order, where
\begin{align}\label{toploop88}
\Delta m_H^2 \sim -\frac{N_c}{8\pi^2}y_t^2M_T^2 =
-\left(88\text{ GeV}\right)^2 \left(\frac{M_T}{450\text{ GeV}}\right)^2,
\end{align}
so $M_T\approx 450$\,GeV is the scale we expect for natural top partners. The current bound results in a ten-times larger $\Delta m_H^2$, which requires a $\Delta_{\rm FT} \sim 10\%$ fine-tuning to get back to the observed $m_H$.

To reduce fine-tuning, one alternative is to have colorless top partners such that the (one-loop) bound could be much weaker since they could be significantly lighter while escaping direct detection. It would be even better if the top partner is a SM singlet, which is known as Neutral Naturalness, like in twin Higgs models \cite{Chacko:2005pe}. However, this alternative is still based on the idea of symmetry and the cancellation between $\Delta m_h^2|_{\text{top}}$ and $\Delta m_h^2|_{\text{top partner}}$. In this work, we would like to explore another alternative which does not require top partners. The idea is to have a strong dependence of the top Yukawa coupling on the energy scale, $y_t=y_t(\mu^2)$, or even directly on the momentum running through the vertex, $y_t=y_t(k^2)$.

While in the former case, the prefactor entering Eqs~\eqref{toploop0} and \eqref{toploop+} would be much smaller\footnote{This is a qualitative estimate, in principle the contribution to the Higgs mass from the top-loop will probe the top Yukawa from low scales until $\Lambda_{\textrm{NP}}$, with more weight given to the higher scales, justifying our use of the Yukawa at high scale $y_t(\Lambda_{\rm NP}^2)$.}, $y_t(\Lambda_{\rm NP}^2)\!\ll\! y_t(M_{\rm EW}^2)\!\approx\!1$, in the latter, more drastic case, Eq.~\eqref{toploop0} becomes
\begin{align}\label{toploopnew}
\Delta m_H^2|_{\text{top}}
&\sim-i\,2N_c \int_{}^{} \frac{d^4k}{(2\pi)^4}\,y_t^2(k^2)\frac{k^2+m_t^2}{(k^2-m_t^2)^2}~.
\end{align}
If $y_t(k^2)$ now drops around $k^2\sim\Lambda_T^2$, the integration gives
\begin{align}\label{toploopL}
\Delta m_H^2|_{\text{top}}
\sim -\frac{3}{8\pi^2}y_t^2 \Lambda_T^2~.
\end{align}
This behavior implies a nontrivial origin of the large top Yukawa coupling that we see at low energies and some new degrees of freedom (but not top partners) significantly below the NP scale where a completion like SUSY or a CHM takes over, i.e. around the scale $\Lambda_T$ in the case discussed in Eq.~\eqref{toploopnew}. 
In the following, we are going to explore these possibilities and discuss how they could emerge in practice. 

In fact, while a strong reduction of the top-Yukawa at the EW scale, $y_t \ll y_t^{\rm SM}$, would be at variance with Higgs (production and decay) measurements at the LHC \cite{ATLAS:2021vrm, CMS:2020gsy, CMS:2020mpn}, a large modification at the NP scale $\Lambda_{\rm NP} \gg M_{\rm EW}$, which is relevant for the HP, remains phenomenologically viable \cite{Goncalves:2018pkt, Goncalves:2020vyn, MammenAbraham:2021ssc, Bittar:2022wgb}. At high energies, the top quark would behave more as the other quarks, and only obtain its large Yukawa coupling in the IR, {\it i.e.}, 
$y_t (M_{\rm EW}^2) \approx y_t^{\rm SM}(M_{\rm EW}^2)$, but $y_t(\Lambda_{\rm NP}^2)\ll y_t^{\rm SM}(M_{\rm EW}^2)$.

\begin{figure}[!t]
\begin{center}
\includegraphics[height=1.6in]{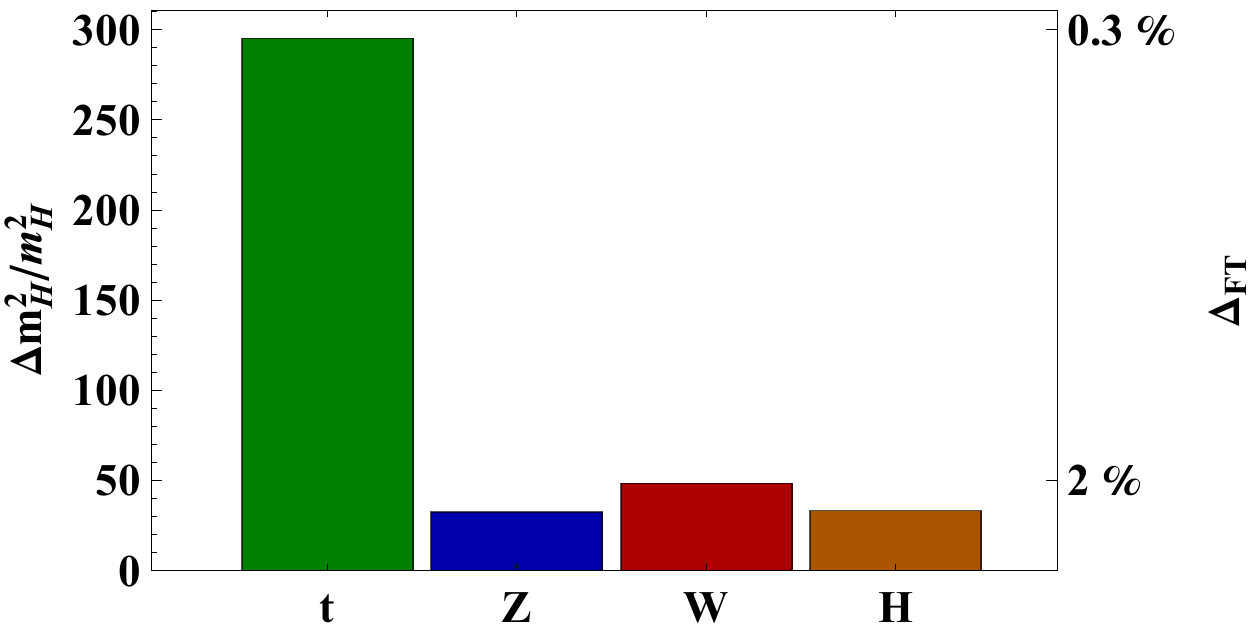}\vspace{-3.4mm}
\caption{\label{fig:DmH1} Various contributions to the Higgs mass squared for $\Lambda_{\rm NP}=10$\,TeV, see text for details. }
\end{center}
\vspace{-3.6mm}
\end{figure}

It is clear from the above discussion, that if $y_t$ would be,
say, a factor of $5$ smaller, the NP coupled to the top could be a factor of $5$ heavier while the
fine tuning $\Delta_{\rm FT}$ would remain the same and the same is true if the top-Yukawa contribution is cut off significantly earlier than other corrections. This is illustrated in Fig.~\ref{fig:DmH1}, where the NP contributions to the Higgs mass term from the 
top-\,, $W$-\,, $Z$-\,, and Higgs sectors, normalized to $m_H$, are presented -- the latter reading \cite{Coleman:1973jx} ($B=W,Z,H$)
\begin{eqnarray}
\label{eq:CWB}
(\Delta m_H^2)_B = \frac 3 2 g_B^2   \frac{\Lambda_{\rm NP}^2}{16 \pi^2}\,,
\end{eqnarray}
with $g_W^2=g^2, g_Z^2=(g^2+g^{\prime 2})/2, g_H^2=4 \lambda$, and we set $\Lambda_{\rm NP}=10$\,TeV.

\begin{figure}[!t]
\begin{center}
\includegraphics[height=1.6in]{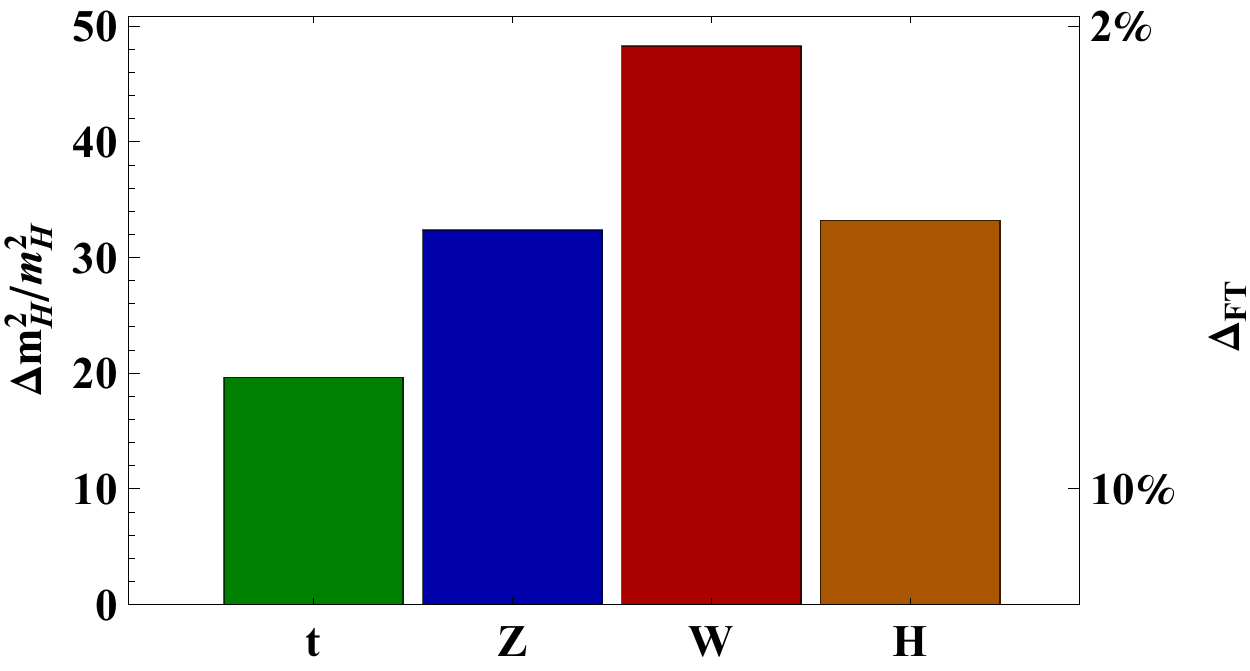}\vspace{-3.4mm}
\caption{\label{fig:DmH2}Various contributions to the Higgs mass squared for $\Lambda_{\rm NP}=10$\,TeV and the top quark Yukawa evolved down due to an additional $U(1)$ gauge boson of mass $M=2.5\,$TeV, with $Y_L=Y_R=2$, $\tilde g_1 = 2.5$, see text for details.}
\end{center}
\vspace{-3.6mm}
\end{figure}

While, by this measure, requiring a tuning of at most 
\begin{equation}
\label{eq:tunCond}
\Delta_{\rm FT} \equiv m_H^2/|\Delta m_H^2| {! \atop \raisebox{3.4mm}{$\gtrsim$}} 5\%
\end{equation} 
would lead to 
\begin{equation}
\Lambda^{\rm SM}_{\rm NP} \lesssim  2.5\,{\rm TeV},
\end{equation} 
the given $\Lambda_{\rm NP}=10$\,TeV results in
$\Delta_{\rm FT}\approx 3 \times 10^{-3}$,
completely driven by the large top contribution,
see Fig.~\ref{fig:DmH1}.
By mitigating the top loop, either via for example $y_t(\Lambda_{\rm NP}^2)\lesssim 0.2 \ll y_t(M_{\rm EW}^2)$ or due to cutting of the top loop at $\Lambda_T \lesssim 1/5\, \Lambda_{\rm NP}$, the situation can be changed significantly. The result is presented in Fig.~\ref{fig:DmH2}, where we see an orthogonal picture to the standard case of Fig.~\ref{fig:DmH1}, namely the top contribution becoming subdominant and the fine tuning remaining modest.

This is made especially relevant by the fact that the first measurement of the running top mass up to
the TeV scale has been provided \cite{CMS:2019jul} which opens an additional handle for testing the idea of significant NP effects in the running top Yukawa, and we will confront the corresponding limits with model predictions
to explore whether this observable could be the first place to see the NP, see also~\cite{Goertz:2016iwa,Alves:2014cda}.

Still, for a full solution to the HP, taming or even cutting completely off the top-loop contribution is not enough and around $\Lambda_{\rm NP}=10$\,TeV, at the latest, a full UV completion -- taking care of contributions from bosonic loops -- should kick in.\footnote{Note that our mechanism becomes even more effective for scenarios where the NP couples mostly through the top sector, leading to further increased cutoff. Moreover, it could also address other new interactions, potentially breaking the Higgs shift symmetry in an energy-scale dependent way. }
The scenario discussed here would thus play an assistant role, relieving the strong bound from the top partner searches, when being combined with traditional solutions like SUSY and CHM. It is especially helpful in composite theories, where an $\mathcal{O}(1)$ Yukawa coupling (breaking explicitly the Goldstone symmetry) is problematic. In fact, this large value is the main reason for the currently biggest tension of such models with
LHC data, which is the absence of expected light top partners \cite{Contino:2006qr,Csaki:2008zd,DeCurtis:2011yx,Matsedonskyi:2012ym,Marzocca:2012zn,Pomarol:2012qf,Carmona:2014iwa} (see also \cite{Blasi:2019jqc,Blasi:2020ktl}).

In the following, we will scrutinize further the microscopic origin of the top Yukawa and the resulting impact on the HP, which could change the conventional picture.  After all, given that the LHC just touches physics beyond the weak scale, it is actually important to ask whether an acceptable fine tuning of $\sim 5\%$ leads to $\Lambda_{\rm NP} \sim 500$\,GeV or could allow for $\Lambda_{\rm NP} \sim 10\,$TeV. Besides expected signatures of the NP related directly to the top Yukawa -- which we will discuss in more detail below -- a 100 TeV collider (and/or precision machine) would then become the ultimate experiment to probe naturalness instead of the LHC. While it could potentially explore in detail the mechanism behind the top Yukawa, null findings at such a collider would drive the NP scale to regions where also the tuning due to the gauge boson and Higgs sectors becomes unnaturally large and thereby our proposed mechanism would no longer be effective in addressing the HP.

The remainder of this article is organized as follows. In Sec. \ref{sec:Running}, we will first consider scenarios with modified renormalization group evolution (RGE) of the top Yukawa and their impact on the fine-tuning in the Higgs mass. We will provide concrete realizations that could drive $y_t$ to smaller values. SM extensions with additional strong interaction are studied from the perturbative regime to setups with non-perturbative dynamics, including a flat extra dimension. Next, in Sec. \ref{sec:OneLoop}, we will present a strongly coupled model where $y_t$ is generated at one loop. Since it is originated from a dimension-six operator, the top loop is fully cut off around the TeV scale, thereby in fact relieving the HP. Finally, the most important experimental tests are discussed in Sec. \ref{sec:Pheno}, before concluding in Sec. \ref{sec:Conclusion}.

\section{Modified Fine tuning due to Running Top Yukawa} \label{sec:Running}

We recall that, at one loop, the evolution of the top Yukawa coupling with the energy scale $t=\ln \mu$ in the SM reads 
\begin{equation}
\frac{dy_t}{dt}  = \frac{y_t}{16 \pi^2} (\,9/2\, y_t^2 - 8\, g_3^2 -  \,9/4\, g_2^2 - \,17/12\, g_Y^2)\,,
\end{equation}
where $g_{Y,2,3}$ are the $U(1)_Y,SU(2)_L,SU(3)_c$ gauge couplings. Thus, we can anticipate that any additional gauge interaction would lead to a further reduction of the top Yukawa at high energies, amplifying the modest decrease within the SM \cite{Chetyrkin:2012rz, Bednyakov:2012en}.

\subsection{General analysis}
In fact, adding a new $U(1)$ or $SU(N)$ gauge symmetry to the SM, with coupling strength $\tilde g_N$\footnote{In the following analysis, we assume the running of the gauge coupling is small in the region of interest such that we can assume $\tilde g_N$ is constant.}, under which the (left- and right-handed) quarks are charged, with hypercharges $Y_{L,R}^2$ in case of the 
abelian group, the RGE becomes
\cite{Cheng:1973nv,Vaughn:1981qi,Machacek:1983fi}
\begin{equation}\label{eq:ytrunning}
\frac{dy_t(t)}{dt} = \frac{y_t(t)}{16 \pi^2} (\,9/2\, y_t^2(t) - B(N)\, \tilde g_N^2)\,,
\end{equation}
where $B(N)\equiv 3 (N^2 - 1)/N $, while in the abelian case we have $B(1)\equiv 3 (Y_L^2 + Y_R^2)$.

\begin{figure}[!t]
\begin{center}
\includegraphics[height=1.8in]{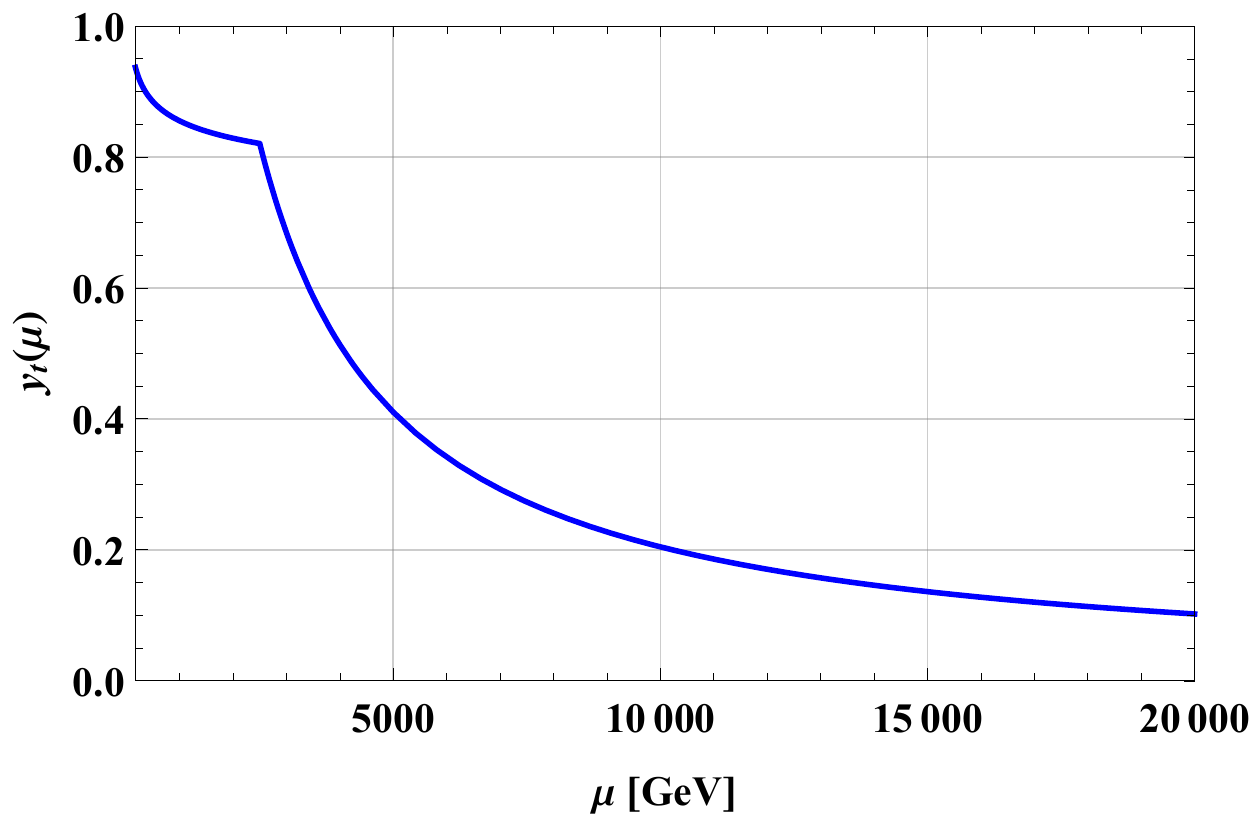}
\caption{\label{fig:run} Running of the top-Yukawa coupling 
in the simple $U(1)$ extension with $Y_L=Y_R=2$, $\tilde g_1 = 2.5$, and a mass of the new gauge boson of $M=2.5\,$TeV, see text for details.}
\end{center}
\end{figure}

Both for simplicity and for getting a first feeling on the potential size of the effect, we focus first on an $U(1)$ extension, where we take $Y_L=Y_R=2$, $\tilde g_1 = 2.5$, and $M=2.5\,$TeV for the mass of the new gauge boson as a benchmark, altering the running above the threshold $M$.
The modified RGE is depicted in Fig.~\ref{fig:run}, where we can inspect that $y_t(\mu\!=\!10\,{\rm TeV}) \approx 0.2$ while $y_t(\mu\!=\!20\,{\rm TeV}) \approx 0.1$, which allows for a significant mitigation of the HP from heavy NP coupled to the top-quark, as we will explore in more detail below.

\subsubsection{Impact on Fine Tuning and $\Lambda_{\rm NP}$}

Recalculating the fine-tuning in the Higgs mass from Eqs.~\eqref{toploop0}/\eqref{toploop+} and \eqref{eq:CWB},
due to new physics residing at the scale $\Lambda_{\rm NP}=10\,$TeV, employing the running top Yukawa coupling $y_t(\mu\!=\!\Lambda_{\rm NP})$ from the $U(1)$ benchmark model above, we arrive at the situation presented in Fig.~\ref{fig:DmH2} and discussed in the introduction. The top contribution becomes subdominant and the fine tuning is driven by the weak bosons, resulting in a rather modest $\Delta_{\rm FT}$. Now, the condition \eqref{eq:tunCond} allows in fact for
\begin{equation}
\Lambda_{\rm NP} \approx 7.5\,{\rm TeV}\,,
\end{equation}
beyond the generic LHC reach.

\subsection{Simple perturbative extensions}

We now discuss a few more concrete setups that can impact the running of $y_t$. The crucial working assumption is that the corresponding new particles could be rather light, modifying the RGE from rather low scales without contributing with large thresholds to the Higgs mass, while in turn other new physics (like SUSY or other top partner models) could reside at larger scales than conventionally considered (i.e. $\Lambda_{\rm NP}\gtrsim 10$\,TeV), without creating a {\it little} HP.

As discussed, extending the SM gauge group by a $U(1)$ or $SU(N)$ under which the left and right handed quarks are charged, is one of the simplest ways to
affect the running of the top-Yukawa according to \eqref{eq:ytrunning}.
$U(1)$ extensions of the SM are severely constrained by anomaly cancellation \cite{Allanach:2018vjg}. If one wants to be flavor diagonal in the charge assignments or, less restrictively, only have one set of non-trivial charge assignments applied to some of the generations, one is left with the hypercharge symmetry or a hypercharge symmetry restricted to one or two generations. If one minimally extends the SM by right-handed neutrinos $\nu_R$ that are singlet under the SM gauge group, the $B-L$ symmetry is also anomaly canceling. We will consider for each scenario below a gauge boson mass of $2.5$ TeV.

\paragraph{$(B-L)_3$ Scenario} 
Gauging the latter symmetry only for the third generation~\cite{Babu:2017olk} avoids stringent constraints on a universal $B-L$ boson from the LEP data, which would limit the mass over coupling ratio of the corresponding boson to $M_{X}/g_X>18$\,TeV \cite{Carena:2004xs}. The gauge boson $X_\mu$ couples vectorially to the third generation 
\begin{equation}
    {\cal L}\supset g_XX_\mu \Big(\bar{t}\gamma^\mu t+\bar{d}\gamma^\mu d-3\bar{\tau}\gamma^\mu \tau-3\bar{\nu_{\tau}}\gamma^\mu \nu_{\tau}\Big)\,.
\end{equation}
As we will be interested in the large coupling regime where the effect on the running of the Yukawa is most significant, we need to check whether the width of the boson stays within the perturbative regime which we will take to be $\frac{\Gamma(X)}{M_X}\lesssim50\%$. Due to the larger charge, the partial width to leptons is three times as big as the one into quarks (despite the color factor) leading to
\begin{equation}
    \frac{\Gamma(X)}{M_{X}}=\frac{2}{\pi}g_X^2\,,
\end{equation}
which restricts the coupling to $g_X<0.89$. This makes it difficult to have a considerable impact on the running of~$y_t$ in reality, pushing it merely down to $0.74$ at $10$\,TeV (versus $0.77$ in the SM, which shows the limitations of an anomaly free U(1) extension).\footnote{We note that this cannot be improved significantly by gauging multiple $(B-L)_3$ symmetries since the multiplicity also enters the scattering cross sections that are bounded by perturbative unitarity (mostly third-generation lepton scattering). In the end, this prohibits enhancing the running notably via a large multiplicity factor.}

\paragraph{Purely top-philic setup}
The situation could be improved by avoiding the problematically large couplings to leptons, leading us to a purely top-philic boson, governed by
\begin{equation}
    {\cal L}\supset g_XX_\mu (\bar{t}\gamma^\mu t)\,.
\end{equation}
While the (anomaly canceling) UV completion that can approximate such a setup would be more contrived, the partial width of the vector boson is now reduced to
\begin{equation}
    \frac{\Gamma(X)}{M_{X}}=\frac{2}{8\pi}g_X^2\,.
\end{equation}
This makes couplings as large as $g_X \lesssim 2.5$ possible which allows for an evolution of the top Yukawa down to $y_t=0.55$ at 10 TeV, resulting in only a modest reduction in fine-tuning.

\paragraph{Third generation non-abelian models}

Before moving to scenarios where more drastic changes are possible, we comment on the prospects of non-abelian extensions of the SM gauge symmetry, preferably exclusively coupled to third generation fermions to evade LHC constraints. Such an example is a non-universal LR model~\cite{Hayreter:2019dzc} where the third generation right handed fermions are charged under a second $SU(2)_R$ group. The relevant part of the Lagrangian reads

\begin{equation}
    {\cal L}\supset g_R/2 Z^{\prime}_\mu (\bar{b}_R\gamma^\mu b_R-\bar{t}_R\gamma^\mu t_R+\bar{\tau}_R\gamma^\mu \tau_R-\bar{\nu}_R\gamma^\mu \nu_R)\,.
\end{equation}
In addition to a neutral $Z^{\prime}$ boson, we now also need to take into account a $W^{\prime}$, although LHC constraints from production of the  $W^{\prime}$ are less restrictive~\cite{Hayreter:2019dzc}. The widths for both vector bosons read
\begin{equation}
    \frac{\Gamma(Z^{\prime})}{M_{Z^{\prime}}}= \frac{\Gamma(W^{\prime})}{M_{W^{\prime}}}=\frac{1}{12\pi}g_R^2\,,
\end{equation}
which limits the coupling constant to $g_R<4.3$. The effect of such an $SU(2)$ charge on the top Yukawa can be seen from \eqref{eq:ytrunning} by employing $B(N)\, \tilde g_N^2 \to \frac{9}{4}g_R^2$, where the additional factor 1/2 appears since only the right handed top is charged under the new $SU(2)$ group. Also, due to strong constraints from $Z^{\prime}$ decay into tau pairs, the lower limit on the mass is $m_{Z^{\prime}}>2.5$\, TeV for a maximal coupling constant. Overall the maximal possible reduction at 10 TeV is to $y_t=0.52$. To produce the desired running to $y_t=0.2$ at 10 TeV, the coupling would need to be as large as $g_R\sim 8.2$, outside of the perturbative regime.

Another example is a broken $SU(3)$ gauge symmetry, similar to a heavy version of QCD or known as Topcolor \cite{Hill:1991at}, which only affects the third generation quarks with Lagrangian given by
\begin{equation}
    {\cal L}= g_3' G'^A_\mu (
    \bar{q}_L\gamma^\mu T^A q_L
    +\bar{t}_R\gamma^\mu T^A t_R
    +\bar{b}_R\gamma^\mu T^A b_R)\,.
\end{equation}
The coupling of the heavy $SU(3)$ can be as low as $g_3'\lesssim 4.5$ for the desired top Yukawa of $y_t\sim0.2$ at $10$ TeV. The limit from the width of the heavy gauge bosons, or so-called coloron, 
\begin{equation}
\frac{\Gamma(G^{\prime})}{M_{G^{\prime}}}
=\frac{1}{12\pi}{g'_3}^2\,,
\end{equation}
of $g_3'<4.3$, could thus borderline be fulfilled. We remark that this would in fact already bring the corresponding contribution to the fine tuning down to that of the W/Z and Higgs bosons. The HP might thus be fully relieved in corners of the parameter space of narrow width under a perturbative regime.\footnote{The inclusion of a finite width can be taken into account as in~\cite{Crivellin:2022gfu}, which could impact the running.} For a generic solution, however, studying the non-perturbative regime might be more appropriate.

\subsection{Beyond the perturbative bound}\label{sec:Nonpert}

Keeping the gauge coupling well in the perturbative regime in general restricts the running of top Yukawa coupling, which limits its capability to relieve the HP. To drastically modify the Yukawa at the TeV scale, the gauge coupling $g_X$ should be large, which enters the non-perturbative regime.

The direct consequence of non-perturbative dynamics is the formation of bound states. To simplify the analysis, we restrict ourselves to a minimal setup where only $q_L=(t_L,b_L)$ and $t_R$ participate in the strong interaction. The expected bound state is a scalar field with SM Higgs-like quantum number, which is known as top-Higgs $H_t$ \cite{Chivukula:2011ag, Chivukula:2011dg}.

The properties of the top-Higgs can be described by the Nambu-Jona-Lasinio (NJL) model \cite{Nambu:1961tp, Nambu:1961fr}. At a scale $\mu \ll M_X$, the effective Lagrangian for the $H_t$ is
\begin{align}
\mathcal{L}=\,&|\partial H_t|^2-\tilde{M}^2|H_t|^2-\tilde{\lambda}|H_t|^4 -\tilde{y}_t \,\bar{q}_LH_tt_R+\textrm{h.c.}
\end{align}
The coefficients are functions of $\mu$, which are given by 
\begin{align}
&\tilde{M}(\mu)^2=\left(\frac{4\pi}{\sqrt{NC}}\frac{M_X}{g_X}\right)^{2}\left(1-\frac{g_X^2}{g_c^2}+\frac{g_X^2\,\mu^2}{\,g_c^2M_X^2}\right),\nonumber\\
&\tilde{\lambda}(\mu)=\frac{16\pi^2}{NC}~,\quad
\tilde{y}_t(\mu)=\frac{4\pi}{\sqrt{NC}}~,
\end{align}
where $N$ is the number of colors of the new strongly coupled gauge symmetry and $C\equiv$ ln$(M_X^2/\mu^2)$. The Yukawa coupling $\tilde{y}_t$ describes the strong interaction between the bound state and its components. The critical coupling $g_c=\sqrt{8\pi^2/N}$ is another important feature of the strong dynamics, which further separates the non-perturbative regime into two different phases - the unbroken and broken phase.

\subsubsection{The unbroken phase}\label{sec:Unbroken}

Starting with the unbroken phase, where the coupling $g_X$ is beyond the perturbative bound but below the critical coupling $g_c$, we can then derive the mass of $H_t$ as
\begin{equation}
M_{H_t}=\tilde{M}(\mu)\sim\frac{4\pi}{\sqrt{NC}}\frac{M_X}{g_X}~.
\end{equation}
The mass of the top-Higgs is $\sim M_X$, which is the generic scale we might expect for the heavy bound states. Since $M_{H_t}\gg m_H$, the top-Higgs naturally decouples from the SM Higgs sector and the effect due to the mixing through the top loop is also negligible.

Taking $N=1\,(3)$, the critical coupling is given by $g_c \sim ~9\,(5)$. The benchmark values $g_X=5\,(g'_3=4.5)$ for (non-)abelian cases both belong to this category. Meaning a reduction in $y_t$ from $\mathcal{O}(1)$ at the EW scale to ${O}(0.2)$ at the 10 TeV scale through running is possible in reality. The desired gauge interaction enters the non-perturbative regime with unbroken phase. Therefore, the HP due to top loop contribution would be relieved at the cost of some new broad resonances, including the gauge boson $X_\mu$ and the bound state $H_t$, at the TeV scale.

\subsubsection{The broken phase}\label{sec:Broken}

If the coupling $g_X$ is stronger than the critical coupling $g_c=\sqrt{8\pi^2/N}$, the story changes dramatically. The strong dynamics will lead to a nontrivial VEV for $H_t$, which not only triggers EW symmetry breaking (EWSB) but also generates the top mass directly. It was first studied in the Top Quark Condensation model \cite{Miransky:1988xi, Miransky:1989ds, Bardeen:1989ds}. This type of model aims to explain EWSB using the VEV of $H_t$, but has been ruled out due to the prediction of a heavy top quark with $m_t \sim 600$ GeV. In our case, the SM Higgs should be responsible for EWSB and the top quark mass so the VEV of top-Higgs needs to be small.

The value of $\langle H_t\rangle$ can be estimated using the NJL model. Taking $g_X>g_c$, the coefficient of the quadratic term $\tilde{M}$ becomes negative and leads to a nontrivial VEV as
\begin{equation}
\langle H_t\rangle\equiv v_t=\sqrt{\frac{-\tilde{M}^2}{2\tilde{\lambda}}}\sim\frac{M_X}{g_X}\sqrt{\frac{g_X^2}{g_c^2}-1}~.
\end{equation}
The generic scale of $v_t$ is $\sim{M_X}/{g_X}$, the breaking scale of the strongly coupled gauge symmetry, which is unacceptably large. Therefore, some cancellations between $g_X$ and $g_c$ are required to get a small $v_t$ well below $246$ GeV. Besides, the VEV of $H_t$ also generates an additional top quark mass with 
\begin{equation}
\tilde{m}_t=\frac{1}{\sqrt{2}}\tilde{y}_tv_t~.
\end{equation}
If it is greater than that from the SM Higgs, the top mass is basically generated from the dynamical symmetry breaking triggered by its own. The idea has been studied in Topcolor-Assisted Technicolor \cite{Hill:1994hp} so we will focus on a small $v_t$ case.

In contrast to the previously studied case of an unbroken phase, where the bound states are heavy and naturally decouple from the SM, in the broken phase, the bound states are expected to be light. The top-Higgs doublet $H_t$ decomposes into a neutral top-Higgs $h_t$ and top-pions $\pi_t$, where the mass of $h_t$ is given by
\begin{equation}
M_{h_t}=\sqrt{2\tilde{\lambda}_0}\,v_t=2\tilde{m}_t < 350 \text{ GeV}~.
\end{equation}
and the top-pions are massless. For a realistic model, additional breaking terms are required to generate the top-pion mass. Still, such light states with strong couplings to top quarks are severely constrained, as will be discussed in Sec. \ref{sec:Pheno}.

\subsection{Strong running from an extra dimension}

An interesting scenario with in general large impact on the renormalization group running of the SM couplings is given by extra-dimensional setups. Although warped models and Gauge-Higgs-Unification setups are more attractive in general, being dual to composite Higgs models via the AdS/CFT correspondence (see e.g. \cite{Khojali:2017azj} in the context of unification), we focus on universal (flat) extra dimensions \cite{Appelquist:2000nn}. Here, the presence of a tower of Kaluza-Klein (KK) excitations effectively turns the running from logarithmic to power law \cite{Dienes:1998vh, Dienes:1998vg}. 

The corresponding beta function for the running Yukawa coupling was found to be
\begin{equation}
\beta_{y}^{\rm UED}(N_{\rm KK}) = \beta_{y}^{\rm SM} + (N_{\rm KK}-1) \beta_{y}^{({\rm KK})}\,,
\end{equation}
where $N_{\rm KK}=E R$ is the number of active KK modes below the energy scale $E$, with $R$ the length of the extra dimension,
while $\beta_{y}^{({\rm KK})}$ is the universal contribution from a single KK level (see, e.g. \cite{Bhattacharyya:2006ym}).
The presence of the factor $N_{KK}$ is the origin of the power-law running.
Taking $R^{-1} \approx 1\,$TeV, this can also lead to
\begin{equation}
y_t(20\,{\rm TeV}) \sim 0.2\,.
\end{equation}
An even stronger running can be achieved with more extra dimensions. It would be very interesting to experimentally extract the running top Yukawa coupling at even higher scales as a probe of such potential power-law running, which has been discussed in \cite{Goncalves:2018pkt}.

The combination of a new gauge interaction and an extra dimension provides another promising scenario. If we let an additional top-philic gauge boson propagates into the bulk, a tower of KK-mode resonances will then modify the running dramatically, similar to $U(1)^N$.


\section{Top Yukawa coupling generated at one loop} \label{sec:OneLoop}

\begin{figure}[tbp]
\centering
\includegraphics[width=0.5\textwidth]{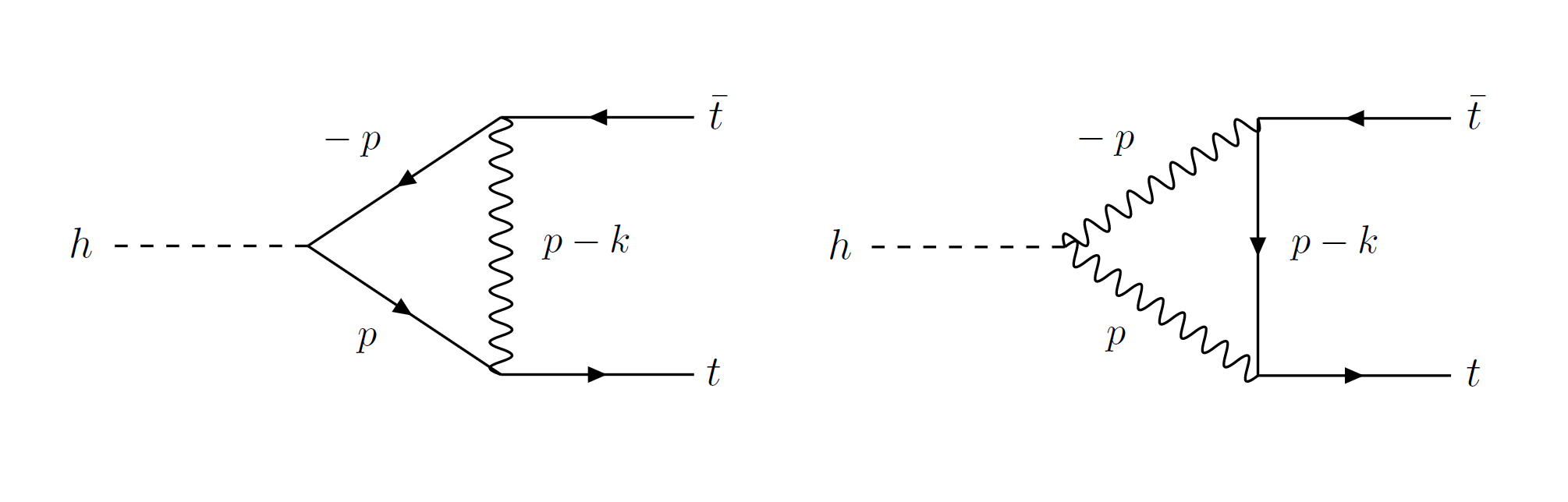}
\caption{Two types of loop diagrams which can generate the top Yukawa coupling. The solid lines represent a fermion and the wavy line could be either a vector (gauge) boson or a scalar boson.\label{vertex}}
\end{figure}

In the previous section, the top Yukawa coupling already exists at high energy but with large running when moving to low energy. In the following discussion, we present another possibility, where the top Yukawa is generated from one-loop effects. Sticking to the elementary Higgs and top quarks at three legs, there are two possible loop diagrams as shown in Fig. \ref{vertex}. The left diagram is the typical top Yukawa-coupling running as we have discussed. 

The right diagram presents a new option in which the top Yukawa coupling is \emph{generated} from a one-loop diagram with the two wavy lines being both scalar bosons or vector bosons, and the solid line representing a new vector-like fermion. The integration of the loop will result in a $1/M^2$ suppression meaning it comes from a dimension-6 operator.

In this case, the top Yukawa coupling does not exist at high energy, but only appears once the heavy degrees of freedom around the scale $M$ are integrated out. The mass $M$ of these new heavy particles play the role of $\Lambda_{\text{T}}$, above which there is no top Yukawa and consequently no top loop correction to $\Delta m_H^2$.

\begin{figure}[tbp]
\centering
\includegraphics[width=0.4\textwidth]{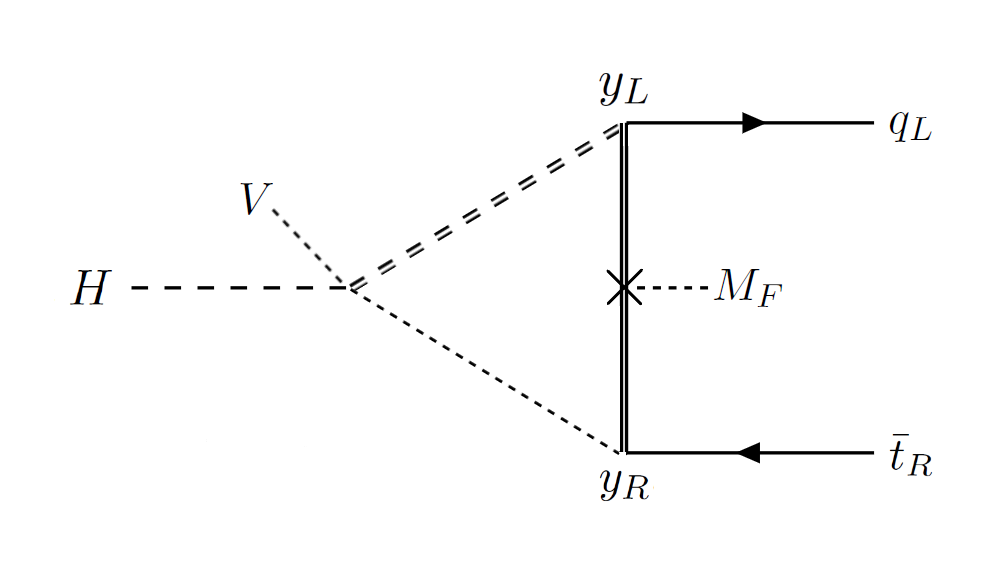}
\caption{Loop diagram which can generate the top Yukawa coupling from a dimension-6 operator. The solid lines represent a fermion and the dashed lines correspond to scalar bosons, see text for details.\label{vertexC}}
\end{figure}

To better understand the diagram, we use the scalar bosons as an example and restore the electroweak symmetry in Fig. \ref{vertexC}. The loop with $1/M^2$ suppression should be compensated by two additional mass scales inserted. One is from the trilinear coupling on the left (i.e. $V$) and the other corresponds to the mass of the vector-like fermion on the right (i.e. $M_F$), which is required to flip the chirality.

At low energies, the particles inside the loop are integrated out, which leads to a $D=6$ operator 
\begin{align}\label{op}
\Delta\mathcal{L}_{y_t}\sim -\frac{1}{16\pi^2}\frac{1}{M^2}(S_V^* S_F)(\bar{q}_L{H}t_R)~,
\end{align}
where $S_V$ and $S_F$ are auxiliary scalar fields corresponding to the $V$ in the trilinear coupling and the vector-like fermion mass $M_F$, respectively. The operator should be responsible for the observed top Yukawa coupling as
\begin{align}\label{yt0}
{y_t}\sim \frac{1}{16\pi^2}y_Ly_R\frac{V\,M_F}{M^2}\sim 1~,
\end{align}
where $M$ represents the heaviest particle in the loop. To realize this idea, the couplings need to be strong enough, since we are trying to generate $y_t\sim1$ from one loop. This might look a bit borderline at first glance, however, we can imagine the diagrams actually originating from some strongly coupled UV theory, then even at one-loop level, the resulting top Yukawa coupling is generically expected to be $\sim 4\pi$. With a strongly coupled theory in mind, what is required is actually an additional suppression to the Yukawa from $4\pi$ down to $1$. More discussions about the possible strongly coupled UV completions are given in the Appendix. In this section, we focus on a simplified model.

\subsection{A simplified scalar model}\label{sec:Model}

\begin{figure}[tbp]
\centering
\includegraphics[width=0.5\textwidth]{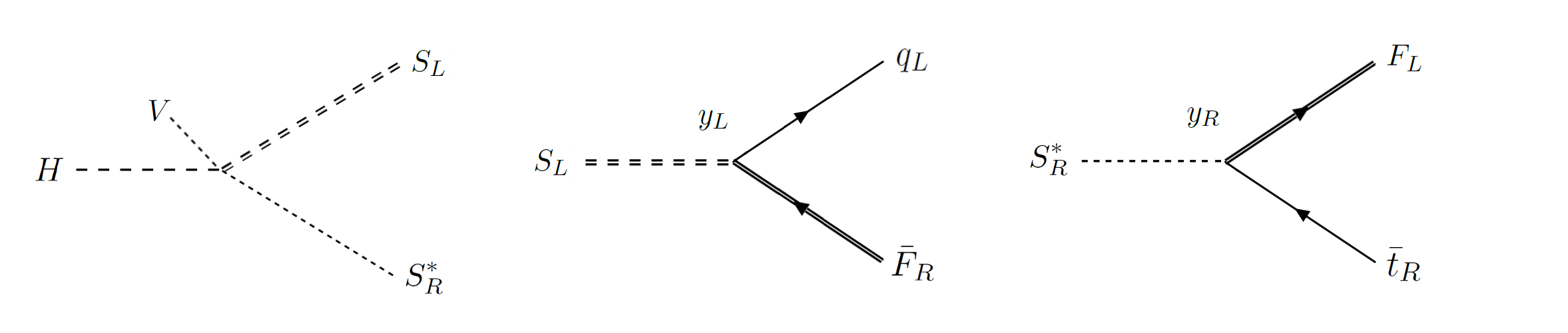}
\caption{Three additional couplings for Higgs (left), the left-handed top quark (center), and the right-handed top quark (right) with scalar bosons and vector-like fermions.\label{coupling3}}
\end{figure}

To generate the top Yukawa coupling, at least three couplings are required as shown in Fig. \ref{coupling3}, which include
\begin{align}\label{required}
\mathcal{L}_{\text{int}}
= -\,VS_{R}S_{L}^\dagger H
- y_{L}\bar{q}_L S_L F_R 
- y_{R}\bar{t}_R S_R F_L
+\textrm{h.c.}~,
\end{align}
where the scalar $S_{L}$ is a doublet and $S_{R}$ and the vector-like fermion $F$ are singlets under $SU(2)_L$. The hypercharge is not determined by the mechanism due to the accidental $U(1)$ symmetry within the particles inside the loop. In general, the hypercharge can be
\begin{align}\label{hypercharge}
Q(F)=Q_F,\quad Q(S_L)=\frac{1}{6}-Q_F,\quad Q(S_R)=\frac{2}{3}-Q_F~.
\end{align}
In the following analysis, we assume $Q_F=2/3$, resulting in the same charge for the vector-like fermion as for the right-handed top quark. $S_L$ then shares the same quantum numbers as the Higgs doublet and $S_R$ becomes a SM singlet scalar. Such an assumption makes it possible to embed the model in a custodial setup, which will be discussed in Sec. \ref{sec:Custodial}. Beside these interactions, we need masses for all the new particles, reading
\begin{align}
\mathcal{L}_{\text{mass}}= - M_L^2|S_L|^2 - M_R^2|S_R|^2 - M_F \bar{F}_L F_R+\textrm{h.c.}~.
\end{align}

\subsubsection{Rotation to mass eigenstates}

To understand the underlying mechanism, we focus on the neutral scalars and move to the mass eigenstates. The relevant Lagrangian terms include
\begin{align}
\mathcal{L}_{\text{neutral}}
&=|\partial S_L|^2+|\partial S_R|^2-M_L^2|S_L|^2-M_R^2|S_R|^2\nonumber\\
&\quad -VS_{R}S_{L}^\dagger H+\textrm{h.c.}\\
&=|\partial s_L|^2+|\partial s_R|^2-M_L^2|s_L|^2-M_R^2|s_R|^2\nonumber\\
&\quad -M_{LR}^2 (s_{L}^*s_{R} + s_{R}^*s_{L})
\end{align}
where $s_{L}$ and $s_{R}$ are the (complex) neutral components of $S_{L}$ and $S_{R}$, and the coefficient of the mixing terms is $M_{LR}^2\equiv V\langle H\rangle=Vv/\sqrt{2}$.

The trilinear coupling leads to mixing between $s_L$ and $s_R$ and therefore we perform a non-trivial rotation to the mass eigenbasis
\begin{equation}
\left( \begin{array}{c} s_L \\ s_R \end{array} \right)
= \begin{pmatrix} c_\beta & -s_\beta \\ s_\beta & c_\beta \end{pmatrix} \left( \begin{array}{c} s_{h} \\ s_{\ell} \end{array} \right)\,,
\end{equation}
where $c_\beta \equiv \text{cos }\beta$, $s_\beta \equiv \text{sin }\beta$, and $s_h$ and $ s_{\ell}$ denote the resulting heavy and light scalars. The angle satisfies the relation
\begin{align}
s_\beta c_\beta &= M_{LR}^2 / \sqrt{4M_{LR}^4+\left(M_L^2-M_R^2\right)^2}.
\end{align}
In the mass basis, the Lagrangian simply becomes
\begin{align}
\mathcal{L}_{\text{neutral}}
=|\partial s_h|^2+|\partial s_\ell|^2-M_s^2|s_h|^2-m_s^2|s_\ell|^2\,,
\end{align}
with $M_s (m_s)$ the mass of the heavy (light) neutral scalar. Their values are given by
\begin{align}
M_s^2(m_s^2)&=\frac{1}{2}\left(M_L^2+M_R^2\right)\pm\sqrt{M_{LR}^4+\frac{1}{4}\left(M_L^2-M_R^2\right)^2}~.
\end{align}

The couplings between the Higgs boson and the scalars in the mass basis become
\begin{align}
\mathcal{L}_{\text{trilinear}}=
&-\sqrt{2}\,Vc_\beta s_\beta \,h|s_h|^2+\sqrt{2}\,Vc_\beta s_\beta \,h|s_\ell|^2\nonumber\\
&-\frac{V(c_\beta^2-s_\beta^2)}{\sqrt{2}}\,hs_{h}^*s_{\ell}+\textrm{h.c.}
\end{align}
and the interaction terms between the scalars and the vector-like fermion read
\begin{align}\label{Lint}
\mathcal{L}_{\text{fermion}}
=&-\left(y_{L}c_\beta \,\bar{t}_L s_h F_R  + y_{R}s_\beta \,\bar{t}_R s_h F_L \right)\nonumber\\
&-\left(- y_{L}s_\beta \,\bar{t}_L s_\ell F_R  + y_{R}c_\beta \,\bar{t}_R s_\ell F_L \right)+\textrm{h.c.}\,.
\end{align}
Having at hand these terms, we can calculate the top Yukawa coupling and related quantities.

Notice that the interactions between the Higgs and new degrees of freedom will also induce additional corrections to $m_H^2$. In this model, trilinear couplings between the Higgs and scalars are introduced, which can form new scalar loops and thus new contributions given by
\begin{align}\label{scalarloop}
\Delta m_H^2|_{\text{scalar}}
\sim \frac{1}{16\pi^2}V^2
\,\text{ln}\left(\frac{\Lambda^2_{\rm NP}}{M^2}\right)~.
\end{align}
This loop is however not quadratically sensitive to $\Lambda_{\rm NP}$ and will thus not reintroduce a HP (note that $V\ll \Lambda_{\rm NP}$). In a later section, we will discuss how the fine tuning condition is affected by the new contribution.

\subsection{Top Yukawa coupling from one loop}

\begin{figure}[tbp]
\centering
\includegraphics[width=0.5\textwidth]{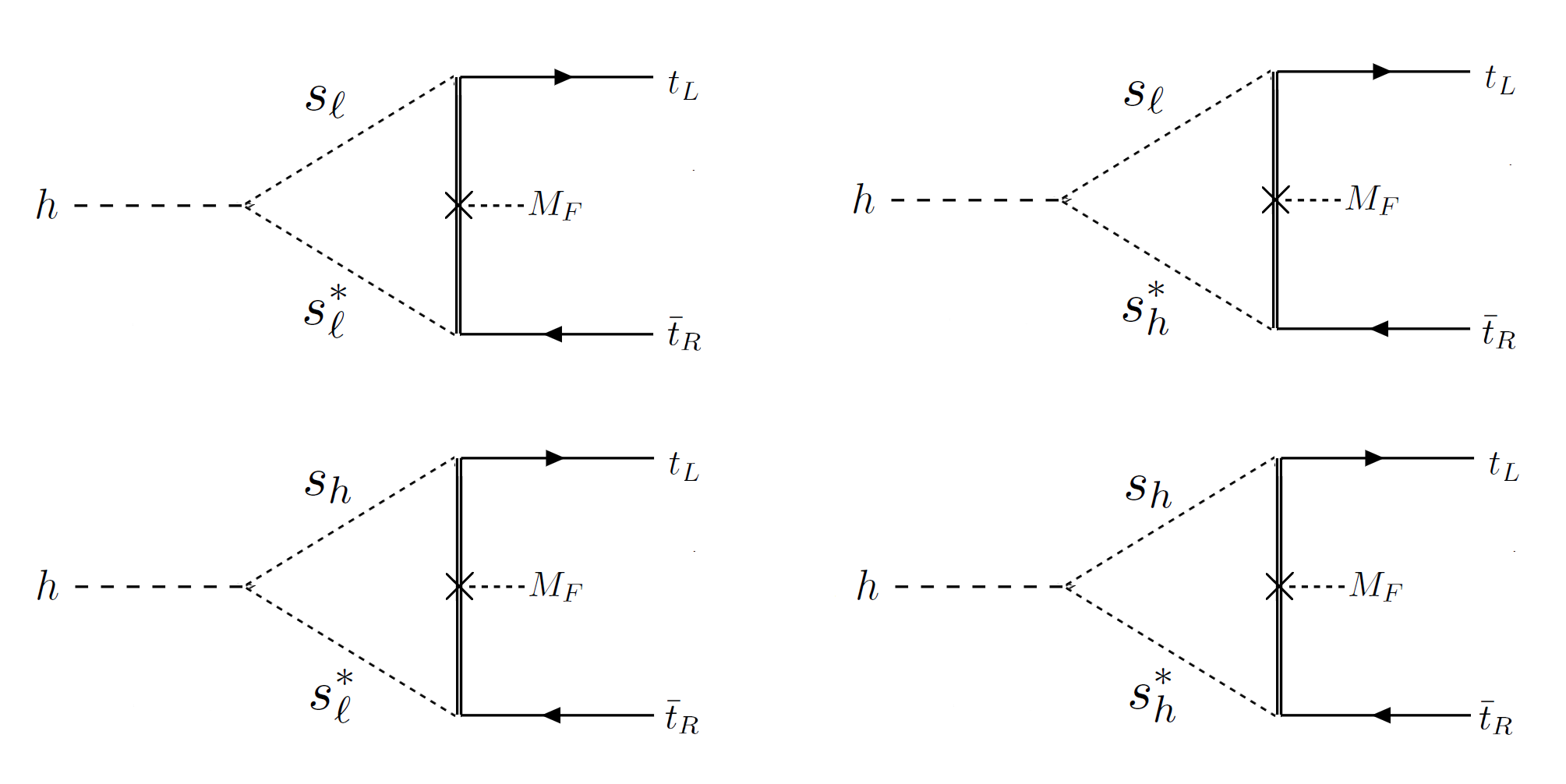}
\caption{The four loop diagrams which contribute to the top Yukawa coupling: Loop 1 (upper left) with two light scalars, Loop 2 (upper right) and Loop 3 (lower left) with both heavy and light scalars, and Loop 4 (lower right) with only heavy scalars.
\label{ytloop}}
\end{figure}

With the interactions described in the last section, we are able to generate the coupling between top quarks and the Higgs doublet by integrating out the new heavy degrees of freedom. At the low energy regime, we get a series of higher dimensional operators in an effective field theory (EFT), reading
\begin{align}\label{expansion}
\mathcal{L}_{\text{top}}
=c_6\left(\bar{q}_L{H}t_R\right)+c_{6+4n}\left(H^\dagger H\right)^n\left(\bar{q}_L{H}t_R\right) ,
\end{align}
where the subscript of the coefficients labels the real dimension of the operators. The first term with coefficient $c_{6}$ corresponds to the SM-like top Yukawa coupling. However, additional higher dimensional operators are also generated. They will result in a deviation of the top-Higgs properties from the SM, which will be discussed later. Moving toward higher energies, the coefficients of the top-Higgs $n-$point interactions will change drastically, replacing the EFT description by form-factor couplings with momentum dependence. For example, in the first interaction, $c_{6}$ will be replaced by a form factor scaling as $\sim 1/(q^{2}-M^2)$ at higher energy and dropping above the scale $M$, where the heavy degrees of freedom can go on-shell.

To derive this behavior explicitly, a loop calculation is required. Since we are interested in the top Yukawa coupling, i.e. the coefficient of $h\bar{t}t$, we work in the $SU(2)$ broken basis. In the mass basis, we have two neutral scalars $s_h$ and $s_\ell$. Therefore, the calculation of $y_t$ actually includes four diagrams as shown in Fig. \ref{ytloop}. The contribution from each diagram is given by
\begin{align}
\text{Loop 1: }&2\,Vy_Ly_R\,c_\beta^2s_\beta^2\int [s_\ell,s_\ell,F]\\
\text{Loop 2: }&Vy_Ly_R(c_\beta^2-s_\beta^2)(-s_\beta^2)\int [s_\ell,s_h,F]\\
\text{Loop 3: }&Vy_Ly_R(c_\beta^2-s_\beta^2)c_\beta^2\int [s_h,s_\ell,F]\\
\text{Loop 4: }&2\,Vy_Ly_R\,c_\beta^2s_\beta^2\int [s_h,s_h,F]\,,
\end{align}
where the square bracket symbolically denotes the triangle-loop integration featuring the respective fields.
Together we obtain an overall contribution of 
\begin{align}\label{ytcal}
y_t=&\,Vy_Ly_R\,\Big(\,(c_\beta^2-s_\beta^2)^2\int [s_\ell,s_h,F]\nonumber\\
&+2\,c_\beta^2s_\beta^2\int [s_\ell,s_\ell,F]+2\,c_\beta^2s_\beta^2\int [s_h,s_h,F]\,\Big)\,.
\end{align}
With $M$ as the heaviest particles in the loop, we get roughly
\begin{align}\label{yt1}
{y_t}\sim Vy_Ly_R\frac{1}{16\pi^2}\frac{M_F}{M^2}~,
\end{align}
which is the same as our estimate in Eq.~\eqref{yt0}.

\subsection{Radiative top mass generation}

\begin{figure}[tbp]
\centering
\includegraphics[width=0.5\textwidth]{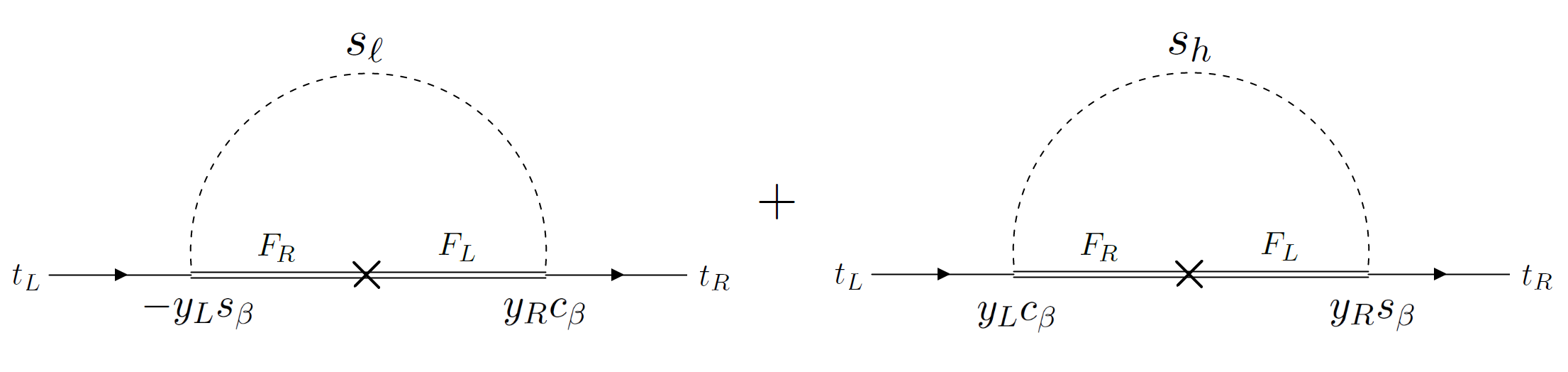}
\caption{The two loop diagrams which contribute to the top mass. Loop 1 (left) with a light scalar and Loop 2 (right) with a heavy scalar.  \label{mtloop}}
\end{figure}

One direct consequence of the model is that now, the top quark mass is generated radiatively from the vector-like fermion $F$. To calculate the generated top mass, two diagrams need to be included as shown in Fig. \ref{mtloop}.

The contribution from the two diagrams is given by
\begin{align}
\text{Loop 1: }&-y_Ly_Rc_\beta s_\beta \int [s_\ell,F]\\
\text{Loop 2: }&y_Ly_Rc_\beta s_\beta \int [s_h,F]~,
\end{align}
Starting from the UV, where $m_t=0$, the top mass is generated via radiative corrections due to the two loops, which gives
\begin{align}\label{mtcal}
m_t=&\,y_Ly_Rc_\beta s_\beta\,\left(\int [s_\ell,F]- \int [s_h,F]\right).
\end{align}
With $M$ as the heaviest particles in the loop, we get
\begin{align}
{m_t}&\sim y_Ly_Rc_\beta s_\beta \frac{M_F}{16\pi^2}\,\left(\frac{M_s^2-m_s^2}{M^2}\right)= \frac{y_Ly_R}{16\pi^2} \frac{M_F V}{M^2}\frac{v}{\sqrt{2}}
\end{align}
which is consistent with the estimation of the top Yukawa coupling in Eq.~\eqref{yt1}. However, notice that the top mass and Yukawa coupling get contributions from all the terms in Eq.~\eqref{expansion}. Therefore, due to the higher dimensional operators, a nontrivial $\kappa_t\equiv y_t/y_t^\text{SM}$ is expected and exact numerical calculations are required.

\subsection{Exact expressions and QCD effects}\label{SEC:Exact expressions}

For the more general case of arbitrary mass hierarchies between the particles $F,s_h$ and $s_l$, one must compute the UV finite loop integral \eqref{mtcal}, which gives a momentum dependent mass
\begin{align}\label{momentumMass}
    m_t(p)=&\frac{y_L y_R c_\beta s_\beta}{16\pi^2} M_F \notag \\ \times &\int_0^1 \textrm{d}x\ln \Big( \frac{p^2 x^2-x p^2+x M_F^2+(1-x)m_s^2}{p^2 x^2-x p^2+x M_F^2+(1-x)M_s^2} \Big)\,.
\end{align}
Similarly, the top Yukawa loop of the light scalar for momenta $p$ and $p^\prime$ of the top quarks is UV finite and reads
\begin{align}\label{momentumYuk}
 &\int [s_\ell,s_\ell,F](p,p^\prime) = \frac{M_F}{32\pi^2}\int_0^1 \textrm{d}x \int_0^{1-x} \textrm{d}y \notag \\
 &\times \frac{2}{p^2x(x-1)+p^{\prime 2}y(y-1) +M_F^2(1-x-y)+m_s^2(x+y)}\,,
\end{align}
to which we should add the analogous heavy scalar loop. The mixed loops will vanish in the maximal mixing scenario, $s_\beta=c_\beta=1/\sqrt{2}$, that we will consider for our benchmarks. Note that both \eqref{momentumMass} and \eqref{momentumYuk} have the following approximate momentum dependence along space-like momentum $p^2=p^{\prime 2}=-Q^2$
\begin{equation}\label{LambdaApprox}
    m_t(Q^2) \sim \frac{m_t(Q^2=0)}{(1+Q^2/\Lambda_m^2)^n}\,,\hspace{1em} y_t(Q^2) \sim \frac{y_t(Q^2=0)}{(1+Q^2/\Lambda_y^2)^n}\,,
\end{equation}
with $n=1$ and where $\Lambda_m\sim \Lambda_y \sim M_F+M_s$. On top of these one-loop effects, we consider one-loop QCD loop effects which can result in large corrections. Indeed on dimensional grounds we expect these to be of order
\begin{equation}
    m_{QCD}(p=0) = \frac{\alpha_s}{\pi} C_F m_t(p=0) \ln(\Lambda^2_m/m_t^2)\,,
\end{equation}
which accounts for corrections of order $25 \%$ and similar for the QCD contribution to the top Yukawa. We note that expressions for non-zero $p,p^\prime$ can be derived and are used in subsequent calculations.

\subsection{Analysis of the Fine Tuning}
Let us consider two benchmarks: one rather conservative (BM1) and one with more striking effects (BM2)
\begin{align}
    M_F&=1530 \textrm{ GeV}, m_s=0.4M_F, M_s = 0.9 M_F \textrm{ (BM1)} \notag \\
     M_F&=865  \textrm{ GeV}, m_s=0.5M_F, M_s = 1.5 M_F \textrm{ (BM2)}\,.
\end{align}
These two benchmarks feature the following effective scales
\begin{align}
    \Lambda_m&=3230 \textrm{ GeV}, \Lambda_y=2980 \textrm{ GeV}, \textrm{ (BM1)} \notag \\
     \Lambda_m&=2220 \textrm{ GeV}, \Lambda_y=1840 \textrm{ GeV}, \textrm{ (BM2)}\,.
\end{align}

The main difference between the benchmarks is that the first includes an experimentally safe deviation from the SM top Yukawa, $\kappa_t=1.1$, while the second one features a larger deviation of $\kappa_t = 1.32$ (which could in principle be brought down by further new physics).

Both are realized with large Yukawa couplings $y_L=y_R=7$ (after inclusion of the QCD loop effects), necessitating a strongly coupled origin, which will be discussed in the Appendix. One can now estimate the level of fine-tuning associated with the top-Yukawa. The usual divergent top loop contribution to the Higgs mass $-\Sigma_h(p=0)$ can be read off from Eq.~\eqref{toploopnew}. When rotating to Euclidean space, the top Yukawa is probed in space-like direction where the expression \eqref{LambdaApprox} is valid, leading to a formally UV finite contribution to the Higgs mass
\begin{equation}
   \Delta m_H^2=-\frac{3 y_t(p=0)^2 \Lambda_y^2}{8\pi^2(2n-1)}\Big[1-(1+(\Lambda_{\textrm{NP}}/\Lambda_y)^2)^{-(2n-1)}\Big].
\end{equation}
For $n>1$, the expression in brackets is not sensitive to the exact value of $\Lambda_{NP}$ and the fine-tuning goes as 
\begin{equation}\label{fine-tuning}
   \Delta_{\textrm{FT},n>1}=\frac{3 y_t(p=0)^2 \Lambda_y^2}{8\pi^2(2n-1)(88 \textrm{ GeV})^2}\,,
\end{equation}
while for $n=1$, the value of where the top-loop is cut off by top partners has an impact on the amount of tuning. At worst, when the top-loop is not cut off ($\Lambda_{NP}\rightarrow\infty$), the upper bound on fine-tuning is finite and is given by the expression above. However when the top-loop is cut off around $\Lambda_{NP}\sim \Lambda_y$ the fine-tuning is halved
\begin{equation}
   \Delta_{\textrm{FT},n=1}=\frac{3 y_t(p=0)^2 \Lambda_y^2}{16\pi^2(88 \textrm{ GeV})^2}\,,
\end{equation}
leading to a fine tuning of $\Delta_{\textrm{FT},n=1}$ of $\sim 5\%$ for both benchmarks. The lower $\Lambda_y$ in BM2 is partly compensated by the higher top Yukawa $y_t(p=0)$.  In the case of a composite top quark consisting of $n$ constituent {\it preons}, the more general formula \eqref{LambdaApprox} for $n>1$ would hold \cite{Lepage:1979za} and could lead to a drastic reduction in fine tuning. 

As mentioned before, it is important to check that the newly introduced interactions do not reintroduce large corrections to the Higgs mass. In our benchmark, scalar loops from two states -- either both heavy or both light -- introduce the corrections 
\begin{equation}
\Delta m_H^2
 = \frac{(V/\sqrt{2})^2}{16\pi^2} (\ln\big(\frac{\Lambda_{\rm UV}^2}{m_s^2}\big)+\ln\big(\frac{\Lambda_{\rm UV}^2}{M_s^2}\big) ) \sim \frac{V^2}{16\pi^2}\,,
\end{equation}
where $\Lambda_{\rm UV}$ is the scale of the strongly coupled UV theory. Assuming a low-scale UV completion, the correction leads to $\sim7\%$ tuning in both benchmarks, which is at the same order as the (reduced) top-quark tuning. Therefore, the new scalar loops do not worsen the fine tuning.

\subsection{Bottom sector and Custodial Symmetry}\label{sec:Custodial}

To generate the bottom Yukawa coupling, we can extend the simplified model in the same way with
\begin{align}\label{bottom}
\Delta\mathcal{L}_{\text{}}
= - M_F^\prime \bar{F'}_L F'_R
- y_L^\prime \bar{q}_L S_L F'_R 
- y'_{R}\bar{b}_R S_R F'_L
+\textrm{h.c.}~,
\end{align}
where the new vector-like fermion $F'$ is a singlet under $SU(2)_L$ with hypercharge $Q(F')=Q_F-1=-1/3$ . While in the following we set $M_F^\prime=M_F$ and $y_L^\prime=y_L$ to respect custodial symmetry. The coupling $y'_R$ is assumed to be much smaller than $y_R$, which violates the custodial symmetry and makes the bottom quark lighter.

Under this setup, we can show that custodial symmetry violation only appears in the subleading order. Indeed, we can rewrite the whole Lagrangian in terms of $SU(2)_L\times SU(2)_R$ representations. The $t_R$ and $b_R$ are combined into a $SU(2)_R$ doublet $q_R$ while the Higgs field is written in matrix form  $\Omega$ as a $(2,2)$ representation. For $Q_F=2/3$, $S_L$ has the same quantum numbers as the Higgs field and can similarly be written in matrix form $\Omega_L \sim (2,2)$. The two vector-like fermions $F$ and $F'$ can be combined to a $SU(2)_R$ doublet $Q_{L/R}=(F_{L/R},F'_{L/R})^T$. Now the whole Lagrangian reads
\begin{align}\label{required}
\Delta\mathcal{L}_{\text{}}
= &-\,VS_{R}\Omega_{L}^\dagger \Omega
- \bar{q}_LY_{L} \Omega_L Q_R 
- \bar{q}_RY_{R} S_R Q_L\nonumber\\
&- M_L^2|\Omega_L|^2 - M_R^2|S_R|^2 - M_F \bar{Q}_L Q_R
+\textrm{h.c.}~,
\end{align}
where $Y_{L}=\text{diag}(y_L,y_L)$ and $Y_{R}=\text{diag}(y_R,y'_R)$ are $2\times2$ matrices. The difference between $y_R$ and $y'_R$ generates the mass splitting between the top and bottom quark and is the only source of custodial symmetry violation in the new physics sector. The related constraints will be further discussed in the next section.


\section{Phenomenology}\label{sec:Pheno}

The modified top Yukawa coupling leaves several imprints in top physics. Since we modify the top Yukawa at the one-loop level, the best test actually comes from indirect measurements, which will be discussed in this section.

\subsection{Running Top mass}\label{sec:mtpheno}

A direct test of the radiative nature of the top Yukawa can be performed in measuring the running of the top quark mass. For the first scenario, the running top mass is affected due to additional heavy gauge bosons, which will shift the curve from the SM prediction around the mass threshold of the heavy bosons. In the second case, the top mass is originated from the loops in Fig. \ref{mtloop} and its running can also be calculated in the same way.

\begin{figure}[tbp]
\centering
\includegraphics[width=0.5\textwidth]{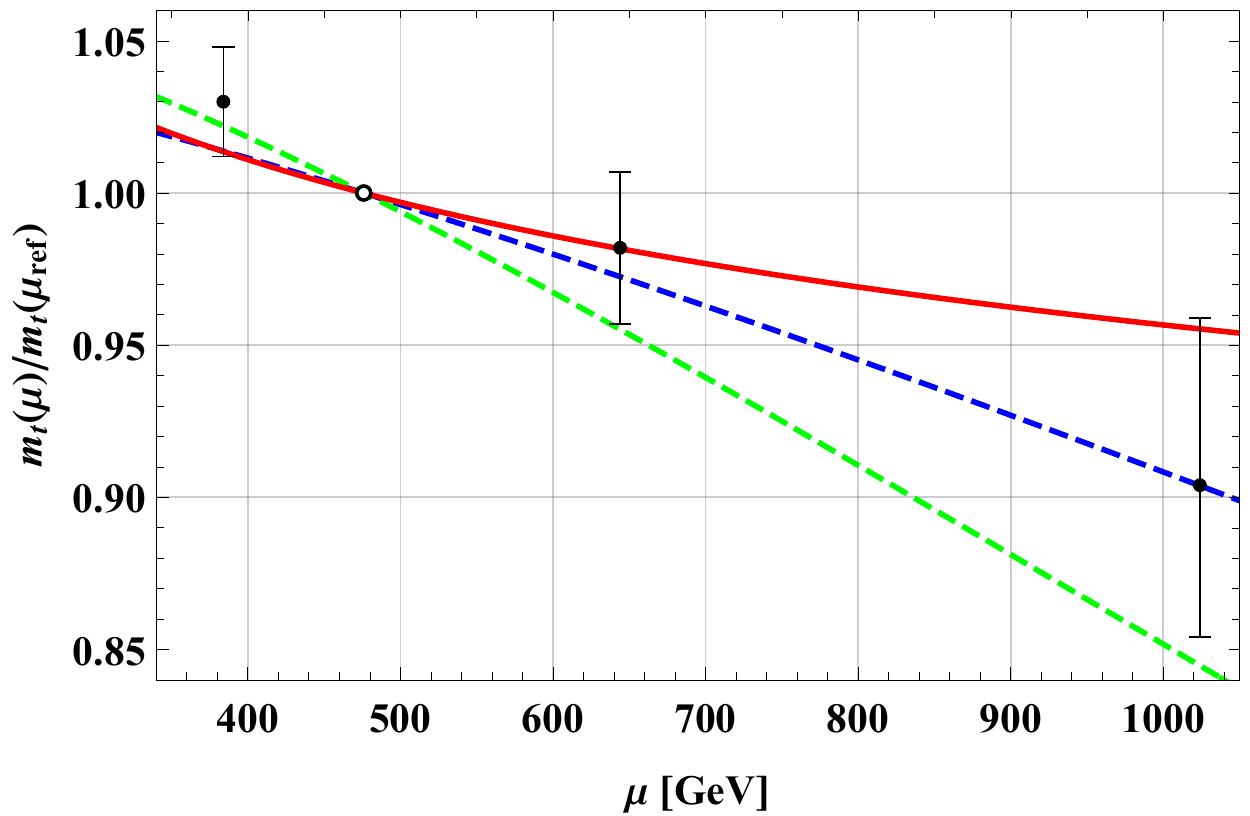}
\caption{The top mass running in the SM (red) versus the running in our conservative BM1 (blue) and for BM2 (green) where the effects are larger, compared with the data points from CMS~\cite{CMS:2019jul}.\label{MassRunning}}
\end{figure}

On the experimental side, the running has been extracted by the CMS collaboration using run 2 data with an integrated luminosity of 35.9 fb$^{-1}$ \cite{CMS:2019jul}. In Fig.~\ref{MassRunning} we confront our results for the two benchmarks with the CMS results. BM2 is already slightly outside of the two-sigma bound in the highest bin, showing the potential of this indirect measurement for such scenarios. The measurement has also been reinterpreted in \cite{Defranchis:2022nqb} but considering the energy scale to be actually half of the original one. The result is then only sensitive up to an energy scale of 0.5 TeV. In this case, the bound on the top mass running becomes weaker and light new physics with less fine tuning is still possible.

\subsection{Top Yukawa coupling measurement}\label{sec:ytpheno}

Another direct test of the idea is the measurement of the top Yukawa coupling. Since we only need a large modification at a high scale to tame the top-loop contribution to the Higgs quadratic term, in general, it is not necessary to have any effect at the EW scale, which is the case for the running top Yukawa scenario. However, in the second scenario, the top Yukawa is generated from the heavy degrees of freedom in the leading order diagrams shown in Fig. \ref{vertexC}. A series of higher dimensional terms as shown in Eq. \eqref{expansion} will also be generated inevitably, which rescales the top Yukawa coupling. The contributions to the top mass from the higher dimensional terms are suppressed by $(V^2v^2/M^4)^n$ ($M$ being the mass of the heaviest particle in the loop), which should be safe with new degrees of freedom being heavy. However, to relieve the HP, we would like to have the new particles as light as possible. Therefore, the bound on $\kappa_t\equiv y_t/y_t^\text{SM}$ will determine how much fine tuning we can relieve.

Both ATLAS and CMS have extracted $\kappa_t$ from the combined measurement of the Higgs boson properties with different production modes and final states. The ATLAS collaboration gets $0.80<\kappa_t<1.04$ at 95\% confidence level \cite{ATLAS:2021vrm}. On the other hand, CMS measures $0.79<\kappa_t<1.23$ with a higher central value but also a larger error bar \cite{CMS:2020gsy}. However, both $\kappa_t$ measurements are closely related to gluon fusion production, which might be modified in a nontrivial way in our model. A more direct measurement from top-associated final states has been done by CMS using $t\bar{t}H$ and $tH$ events. The measurement gives $0.7<\kappa_t<1.1$ at 95\% confidence level \cite{CMS:2020mpn}. But still, with contributions by off-shell top quarks in the processes, the real bound should be weaker.

The $\kappa_t$ in the model can be calculated from the $y_tv/m_t$ using the exact expression from Sec. \ref{SEC:Exact expressions} including QCD effects. $\kappa_t$ is partly driven by the degeneracy of the two scalars (or $V^2/m_S^2)$. Benchmark 1 results in a $\kappa_t=1.1$, while Benchmark 2 results in $\kappa_t=1.32$. Interestingly these types of models tend to increase $\kappa_t$, which tends to increase the amount of fine-tuning. It would be interesting to see if one could obtain a similar type of model with generically $\kappa_t<1$. Although the latter benchmark is quite optimistic, the $y_t$ is quickly driven down which can impact the effective gluon fusion vertex. Indeed, due to the form \eqref{LambdaApprox} of the top Yukawa, one expects this operator to be diminished by $(1-2m_t^2/\Lambda_y^2 \ln(\Lambda_y^2/m_t^2))$, which would transform the $\kappa_t=1.32$ into effectively $\kappa_t=1.21$ which is within the CMS bound.

\subsection{Form factors}\label{sec:ffpheno}

\begin{figure}[tbp]
\centering
\includegraphics[width=0.5\textwidth]{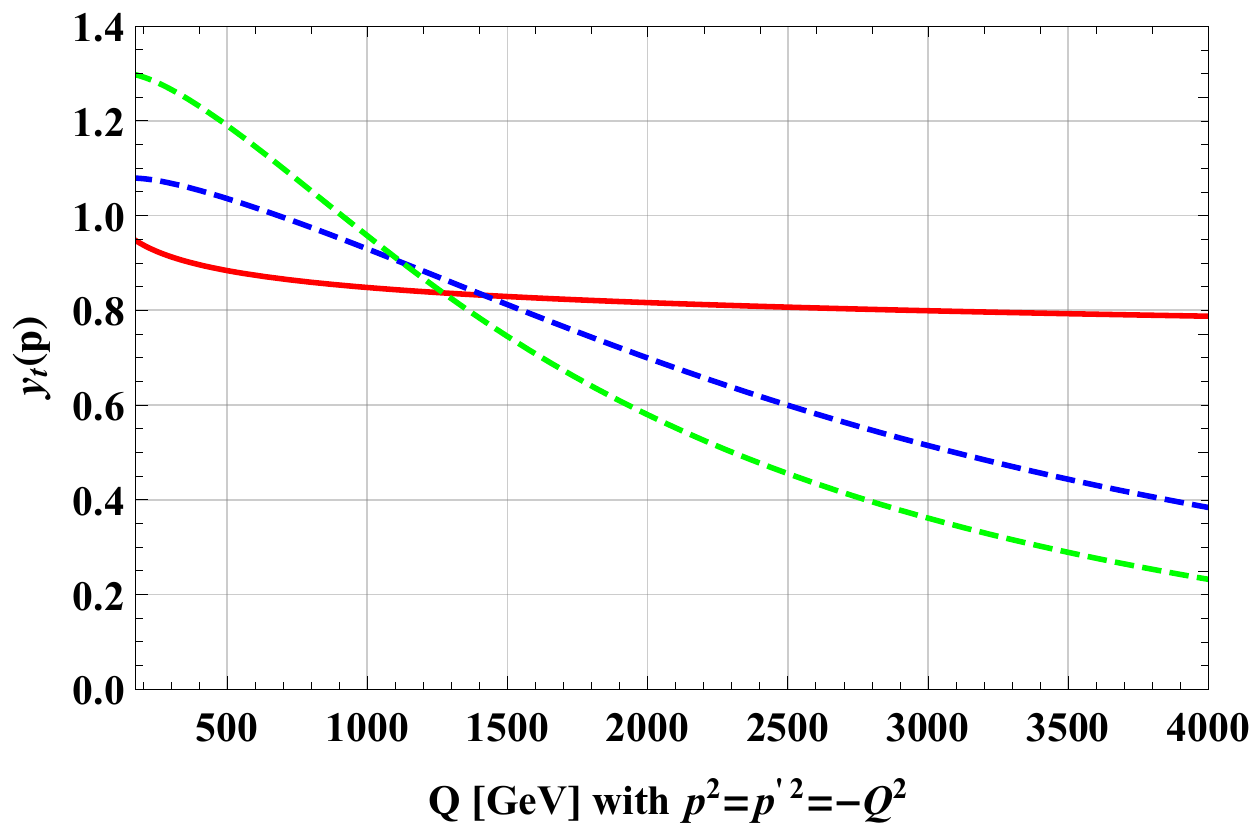}
\caption{The top Yukawa form factor running in the SM (red) versus the running in our conservative BM1 (blue) and for BM2 (green) where the effects are larger.\label{YukawaRunning}}
\end{figure}

To directly probe the top Yukawa coupling beyond the EW scale, we need to derive the top-Higgs form factors, which describe the momentum-dependence of the top Yukawa coupling \cite{Goncalves:2018pkt, Goncalves:2020vyn, MammenAbraham:2021ssc, Bittar:2022wgb}. We present in Fig. \ref{YukawaRunning} the top Yukawa form factor in our two benchmarks, compared to the SM Yukawa form factor. These deviations are especially useful when the off-shellness of the top is comparable to $\Lambda_y$ and a large reduction with respect to the on-shell value is observed. These effects could be seen in the tails of momentum distributions. In processes such as $t\bar{t}h$ production, where the top quark is probed in time-like momentum and thus the resonant pole structure is probed, this could lead to significant increases in the total cross section ~\cite{Bittar:2022wgb} and especially in differential momentum distributions. However a more quantitative statement for the model requires a dedicated analysis which is outside the scope of this work.

\subsection{Four top quarks cross section}\label{sec:4tpheno}

To modify the top Yukawa at the loop level, strong interactions among top quarks are always required. Due to the strongly coupled nature, normal resonance searches are hard to be performed. However, the effect can still be caught in the measurement of the four top-quark cross section. Unlike the resonance searches, this search is more like a precision test due to its small rate. In the SM, the prediction for the cross section is \cite{Frederix:2017wme}
\begin{align}
\sigma_{t\bar{t}t\bar{t}}^{\text{SM}}=12\pm 2.4 \text{ fb.}
\end{align}

Measurements using different final states have been performed by both ATLAS \cite{ATLAS:2020hpj, ATLAS:2021kqb} and CMS \cite{CMS:2019rvj, CMS:2022uga} with LHC run 2 data. The cross section measured by ATLAS is
\begin{align}
\sigma_{t\bar{t}t\bar{t}}^{\text{ATLAS}}=24^{+7}_{-6} \text{ fb,}
\end{align}
whose central value is about two times the SM prediction while CMS gets a value closer to the SM prediction
\begin{align}
\sigma_{t\bar{t}t\bar{t}}^{\text{CMS}}=17^{+5}_{-5} \text{ fb.}
\end{align}
Both collaborations have seen evidence for the simultaneous production of four top quarks and a slightly larger cross section compared to the SM prediction. Although the sensitivity does not yet allow one to claim the existence of four top final states, it is already enough to constrain BSM models with a modified top sector. The bound on the cross section at 95\% C.L. is given by
\begin{align}
\sigma_{t\bar{t}t\bar{t}}^{\text{}}<38\,(27) \text{ fb   from ATLAS (CMS).}
\end{align}

Several analyses aiming at interpreting the results in terms of simplified models or effective field theories have been performed in recent years \cite{Darme:2021gtt, Banelli:2020iau, Blekman:2022jag}. Following the analysis of simplified models in \cite{Darme:2021gtt}\footnote{Notice that the analysis might not be reliable for broad top-philic particles due to the complicated interference with the SM background. However, the detailed analysis of the large-width effect is beyond the scope of this study. In this paper, we assume the original analysis in \cite{Darme:2021gtt} can be extended to a larger width condition.}, we get a constraint on a top-philic vector singlet boson with coupling $g_V$ and mass $M_V$ of 
\begin{align}
\frac{g_V}{M_V}<2.1\,(1.8) \text{   from ATLAS (CMS).}
\end{align}
 at 95\% C.L. Similarly, the bound on a top-philic scalar singlet boson with coupling $g_S$ and mass $M_S$ is given by 
\begin{align}
\frac{g_S}{M_S}<3.0\,(2.6) \text{   from ATLAS (CMS).}
\end{align}
The first constraint can be directly applied on the top-philic vector boson $X_\mu$ for the running top Yukawa scenario with strongly coupled $U(1)$ gauge interaction. The benchmark point in Sec. \ref{sec:Running} gives $g_X/M_X=2$, which is right around the 95\% C.L. bound. On the other hand, the bound on a scalar field is important for the top-Higgs $H_t$, the bound state of top quarks described in Sec. \ref{sec:Nonpert}, especially in the broken phase where the bound states are expected to be light. For the heavy QCD case, the top-philic vector boson, or coloron, is a color octet $G'$, which couples to gluons directly. The pair production of $G'$ will lead to a large cross section in the four top final state for light $M_{G'}<2$ TeV. To get the desired running, $M_{G'}=2.5$ TeV is chosen, where the pair production is subleading. We can here derive the upper bound as 
\begin{align}
\frac{g_3'}{M_{G'}}<2.9\,(2.5) \text{   from ATLAS (CMS),}
\end{align}
which is larger then the benchmark with ${g_3'}/{M_{G'}}\sim 1.7$.

For the loop-generated top Yukawa scenario, the condition is more complicated. Four-top operators are generated at the one-loop level with both scalar-like and vector-like operators including
\begin{align}
a(\bar{t} t)(\bar{t} t),\, b (\bar{t} \gamma^\mu t)(\bar{t} \gamma_\mu t),\, c(\bar{t}\gamma^5 t)(\bar{t}\gamma^5 t),\, d (\bar{t} \gamma^\mu \gamma^5 t)(\bar{t} \gamma_\mu \gamma^5 t).
\end{align}
For the benchmarks, we are able to calculate the coefficients of these operators. For BM1, we get
\begin{align}
(a,b,c,d)=\frac{1}{M_F^2}(1.81,-0.49,-1.40,-0.45),
\end{align}
where $M_F=1530$ GeV. For BM2, the coefficients are
\begin{align}
(a,b,c,d)=\frac{1}{M_F^2}(1.18,-0.37,-0.66,-0.31 ),
\end{align}
where $M_F=865$ GeV. To compare with the experimental constraints, we sum over the operators and rewrite them in the standard basis \cite{Darme:2021gtt} as
\begin{align}
(b+d) O^1_{QQ} + (b+d) O^1_{tt} +  (c-a)/3\, O^1_{Qt} + 2 (c-a) O^8_{Qt},
\end{align}
where the operators with coefficients $(a+c)$ and $(b-d)$ are not listed due to an approximate cancellation among the coefficients which makes the effects subleading. In general (and for our benchmarks in particular) the coefficients of the $O^1_{QQ}$, $O^1_{tt}$ and $O^1_{Qt}$ operators are approximately the same, which allows us to combine them. The combination turns out to be similar to the generated operators from a top-philic singlet vector $V$ with the ratio
\begin{align}
\frac{g_V}{M_V}\sim \sqrt{-2(b+d)},
\end{align}
which for our benchmarks result in $\sim 0.9$ and $\sim 1.4$ for BM1 and BM2 respectively -- both are well below the current constraint. By modeling our NP effects with a top-philic singlet vector, the $O^1_{tt}$ operator is slightly underestimated while the $O^8_{Qt}$ operator is neglected. Although the latter has a larger coefficient, its importance in 4t production is suppressed in comparison to the other operators as EFT analysis shows \cite{Blekman:2022jag}. Therefore, the four top quarks final states can only give a weak constraint on the models with a loop-generated top Yukawa.

\subsection{Flavor constraints}\label{sec:Flavor}

Besides the four top cross section, the same four-quark operators will also have an impact on light quark physics through mixing, which can introduce dangerous flavor changing neutral currents. Assuming that the angle $\theta_{23}\gg\theta_{13}$, analogous to the CKM matrix, then among all the processes, the strongest constraint is expected from $B_s-\bar{B}_s$ mixing, which contains both the second and third generation quarks. The relevant contribution is induced by the operator
\begin{equation}
\Delta\mathcal{L}_{B_s}=C_{sb}(\bar{s}_L\gamma_\mu b_L)(\bar{s}_L\gamma_\mu b_L).
\end{equation}
Following the calculation in \cite{DiLuzio:2017fdq}, we can derive the deviation in the mass difference $\Delta M_s$ as
\begin{equation}
\frac{\Delta M_s}{\Delta M_s^{SM}}
\approx 1+\left(22\,304 \text{ TeV}^2\right)\,C_{sb}^2\,,
\end{equation}
assuming the new physics are around the TeV scale.

The measurement of the mixing parameter \cite{HFLAV:2019otj} compared with the SM prediction by sum rule calculations \cite{King:2019lal}, leads to a bound at 95\% C.L. of
\begin{equation}\label{Bsmixing}
\left| C_{sb} \right| \leq \left(\frac{1}{274~\text{TeV}}\right)^2.
\end{equation}

For the simplest top-philic singlet vector boson $V$, the coefficient is given by
\begin{equation}
C_{sb}\approx -\frac{1}{2}\frac{g_{V}^2}{M_{V}^2} \theta_{sb}^2
~\implies~
\frac{g_{V}}{M_{V}} \theta_{sb} \leq \frac{1}{194~\text{TeV}}~,
\end{equation}
where the angle $\theta_{sb}$ parametrizes the rotation between the second and third generations of down-type quarks in the mass basis. For the benchmark point with $g_X/M_X=2$, the angle $\theta_{sb}$ needs to be smaller than $0.003$. For the second scenario, with smaller four top operators, the constraint is similarly weaker, becoming $\theta_{sb}\lesssim 0.005$.

\subsection{Electroweak precision tests}\label{sec:EWPT}

Treating the top and bottom quarks differently will generically lead to custodial symmetry breaking and additional contributions to the $T$ parameter \cite{Peskin:1990zt,Peskin:1991sw}, which is well measured in electroweak precision tests.

Focusing on the constraint of the $T$ parameter, many analyses on the oblique parameters have been conducted recently \cite{Lu:2022bgw, Strumia:2022qkt, deBlas:2022hdk, Bagnaschi:2022whn, Asadi:2022xiy}. We follow the $S-T$ contour derived in \cite{Asadi:2022xiy}.
At 95\% C.L., one obtains $T\lesssim 0.25$. However, the new $M_W$ measured by the CDF collaboration \cite{CDF:2022hxs} shows a deviation from the SM prediction, which would imply some amount of custodial symmetry violation. Using only the CDF $M_W$ measurement, the $2\sigma$ preferred region would read $0.12\lesssim T \lesssim 0.42$ according to different $S$. Thus, a reasonable benchmark for the $T$ parameter is $\Delta T= 0.25$, which corresponds to the upper bound without the CDF $M_W$ measurement and the central value with only the CDF measurement.

The value can be transformed to the coefficient of the custodial symmetry violating operator
\begin{align}
\mathcal{L}_{T}=c_T\left|H^\dagger D_\mu H\right|^2,\text{ where }\ \Delta T=-\frac{v^2}{2\alpha}c_T~.
\end{align}
The requirement of $\Delta T= 0.25$ becomes
\begin{align}
|c_T|=1/(3.95 \text{ TeV})^{2}~.
\end{align}

For the running top Yukawa scenario, the operator is induced by the top loop with additional strong interaction inside, which leads to 
\begin{align}
c_T\sim c_N\left(\frac{1}{16\pi^2}\right)^2y_t^4g_X^2\frac{1}{M_X^2}~,
\end{align}
with $c_1=3$ and $c_{N>1}=(N^2-1)/2$ for $SU(N)$ vector bosons. The two loop nature suppresses the contribution well below the bound to $|c_T|\sim 1/(45 \text{ TeV})^{2}$ for both the Abelian benchmark ($g_X=5\,,~M_X=2.5$ TeV) and the non-Abelian benchmark ($g'_3=4.3\,,~M_{G'}=2.5$ TeV).

For the loop-generated top Yukawa scenario, a dangerous custodial symmetry violating operator can be directly generated at one-loop with scalars. With an additional custodial symmetry setup presented in Sec. \ref{sec:Custodial}, the scalar sector is protected and the only source of custodial symmetry violation comes from the coupling $Y_{R}=\text{diag}(y_R,y'_R)$. Therefore, the custodial symmetry violating operator is only generated from diagrams with an additional $F-t_R$ fermion loop inserted, which first appears at three-loop level. The two additional loops give an extra suppression factor of $(y_R^2/16\pi)^2$ in comparison to the original one-loop with only scalars and the resulting coefficient is then given by
\begin{align}
c_T\sim \left(\frac{1}{16\pi^2}\right)^3 N_c V^4y_R^4\frac{1}{M_{S/F}^6},
\end{align}
with the color factor $N_c=3$ and $M_{S/F}$ the mass of the heaviest particle in the loop. We find $|c_T|\sim 1/(17 \text{ TeV})^{2}$ for BM1. We also get a larger contribution $|c_T|\sim 1/(3.3 \text{ TeV})^{2}$ for BM2 with light new particles but it is around the same order as the experimental constraint.

\subsection{$Zb\bar{b}$ coupling}\label{sec:Zbb}

Another strong constraint on top quark related models comes from the $Zb\bar{b}$ coupling. Especially, the deviation $\delta g_{b_L}$ from the current experimental central value is constrained within $5\times 10^{-3}$~\cite{Batell:2012ca, Guadagnoli:2013mru} at $95\%$ C.L., while the constraint on $\delta g_{b_R}$ is much weaker at $3\times 10^{-2}$. Considering $|g_{b_L}|\sim|g_{b_R}|$, a negative $g_{b_L}$ with $|g_{b_L}| < 3\times 10^{-3}$ is preferred.

The value can also be transformed into the coefficient of the higher-dimensional operator
\begin{align}
\mathcal{L}_{Zbb}=c_{b}\left(H^\dagger D_\mu H\right)\left(\bar{q}_L \gamma^\mu q_L\right),\text{ where }\ \delta g_{b_L}=-\frac{v^2}{2}c_b~.
\end{align}
The requirement $|\delta g_{b_L}|< 3\times 10^{-3}$ becomes
\begin{align}
|c_b| < 1/( 3.17\text{ TeV})^{2}~,
\end{align}
which is of the same order as the bound on $c_T$.

For the running top Yukawa scenario, the operator originates from a loop with top quarks and the top-philic boson, which results in 
\begin{align}
c_b\sim \frac{c_N}{16\pi^2}y_t^2g_X^2\frac{1}{M_X^2}~,
\end{align}
with $c_1=1$ and $c_{N>1}=(N^2-1)/2N$. The contribution is again suppressed, but somewhat larger compared to $c_T$, reading $|c_b|\sim 1/(5.4 \text{ TeV})^{2}$ for both the Abelian and non-Abelian case.

For the loop-generated scenario with the additional custodial setup, the operator again arises at the three-loop level via the $y_R$ coupling, which gives
\begin{align}
c_b\sim \left(\frac{1}{16\pi^2}\right)^3V^2y_R^4y_L^2\frac{1}{M_{S/F}^4}~.
\end{align}
We find that both $|c_b|\sim 1/(12 \text{ TeV})^{2}$ for BM1 and $|c_b|\sim 1/(4 \text{ TeV})^{2}$ for BM2 are within the experimental constraint.

\subsection{Direct searches}\label{sec:Direct}

Direct searches usually provide the most important tests for models with TeV-scale new degrees of freedom. However, in our case, due to the strongly coupled and one loop nature, direct searches are not that useful.

For the top Yukawa with large running due to additional gauge bosons, the gauge boson as well as the possible formation of bound states both have large widths ($\gtrsim 50\%$), which requires new analyses as opposed to the traditional narrow resonance searches. Also, for the minimal setup, where the gauge boson only couples to top quarks, the final state is exactly four top quarks as we have discussed and the resonance searches do not perform better than the cross section measurement.

For the top Yukawa generated from one loop, new scalars and fermions are introduced, which should lead to exotic phenomenology. However, since the only requirement for the new particles is to form the desired loop, the quantum numbers of these particles are not fixed by the mechanism. This means there are many possibilities for the new states and the corresponding phenomenology. We focus on the most intuitive case, where the vector-like fermion has the same quantum numbers as the right-handed top quark and only the neutral light scalar is relevant. Production of $F\bar{F}$ pair is expected to be the most relevant one. Each $F(\bar{F})$ will then decay to $t(\bar{t})$ with a light scalar $s_\ell$. If the light scalar is stable, then the final states will be $t\bar{t}$ plus missing energy, which is similar to the searches for top squark pair production with stable neutralinos \cite{ATLAS:2020dsf, ATLAS:2021hza, ATLAS:2020xzu, CMS:2019ysk, CMS:2021eha, CMS:2021beq}. The current results exclude a top squark mass up to $1200$ GeV for a $600$ GeV neutralino, which is still far from the BM1 with $M_F=1530$ GeV and $m_s=612$ GeV. However, for BM2 with $M_F=865$ GeV and $m_s=433$ GeV, it is completely excluded. The bound can be avoided, however, with an additional operator involving the light scalar. For example, if it couples to gluons through a new operator $s_\ell GG$, then the final states will become $t\bar{t}$ plus jets, where the background is much larger. A similar search has been conducted by the CMS collaboration \cite{CMS:2021knz}, whose results can be reinterpreted to this model as $M_F>670$ GeV when $m_s=100$ GeV, which is much weaker.


\section{Conclusions} \label{sec:Conclusion}
The hierarchy problem remains a driving force for looking for TeV-scale new physics. The quadratic divergent corrections to the Higgs squared term implies that there should be new degrees of freedom for each Higgs coupling at the TeV scale. Among them, the top quark plays the most important role due to the large top Yukawa coupling. The traditional idea to deal with the top loop is introducing top partners, usually based on some symmetry. The top loop is canceled by the top partner loop such that the quadratic sensitivity to the new scale vanishes. In this study, we presented alternatives to this traditional idea. We discussed scenarios where the top Yukawa coupling drops dramatically at higher scales, such that the top loop contribution is no longer the dominant concern and the new physics scale is allowed to be higher. To realize this idea, we proposed two scenarios to modify the top Yukawa coupling at the loop level, running top Yukawa and loop-generated top Yukawa.

In running top-Yukawa models, additional gauge bosons coupled to top quarks are introduced at the (low) TeV scale and we discussed several types of gauge symmetry. To relieve the hierarchy problem, the desired RGE requires strong gauge coupling. We found that the coupling within the perturbative regime can barely satisfy the demands. For stronger coupling, the bound state $H_t = \bar{t}_Rq_L$ forms, which can be described by the NJL model. The coupling of interest falls in the non-perturbative regime with the unbroken phase, where the bound state is heavy and decouples from the SM particles.

For the second scenario, we generated the top Yukawa at one-loop level. Due to its dimension-six nature, the top loop will be cut off at the characteristic scale and only gives a finite contribution to the Higgs mass. To realize the idea, new degrees of freedom have been introduced, including top-philic scalars and a vector-like fermion. A simplified model as well as two benchmarks were discussed in detail. We showed the finetuning due to the top loop contribution and the additional scalar loop can be controlled to be at the $\sim 5\%$ level.

These alternatives also suggest looking for effects in different places in comparison to the top partner solutions. For the running top Yukawa scenario -- since the new gauge boson needs to directly and strongly couple to the top quarks -- the strongest constraints come from the four-top cross section. Strong coupling is only possible with a large mass, which limits its capability to relieve the hierarchy problem. On the other hand, in the loop-generated top Yukawa model, the four-top operators are likewise loop-generated, which is less constrained. However, there are unavoidable higher dimensional operators, which generate the top mass and top Yukawa in different manners. Thus, the strongest bound comes from the direct top Yukawa measurement. Also, due to the rather light particles allowed in the model, the top mass running starts from the sub-TeV scale, which provides a unique test of the idea, as shown in Fig. \ref{MassRunning}.

The presented modified-top-Yukawa model can assist traditional models like SUSY and CHM. If colored top partners remain unobserved in the future, the mechanism provides a good alternative to relieve the large top loop contribution. Moreover, it points to some distinct signatures such as four-top final states and a running top mass, which are rarely discussed. Especially the running top mass measurement, which directly probes the top Yukawa coupling at higher scales, might unveil the nature of the largest Yukawa coupling in the Standard Model.

\section*{Acknowledgments}

We thank Hsin-Chia Cheng, Gian Giudice, Manfred Lindner, Da Liu, and Markus Luty for useful discussions and comments and Simone Blasi and Rakhi Mahbubani for collaboration in the early stages of this work.

\appendix

\section{A Strongly Coupled UV Theory}\label{sec:UV}

Large couplings are required in the Simplified Scalar Model discussed in Sec. \ref{sec:Model}, which implies a strongly coupled UV-completion. In this Appendix, we present a possible UV theory for the simplified model, which provides a validation of our idea.

\subsection{An $SU(3)_L\times SU(2)_R$ global symmetry}

Before we discuss the strongly coupled theory, we first explore the ingredients it needs to include and the possible symmetry structure behind it. In Eq. \eqref{required}, we show that the minimal setup for generating the top Yukawa coupling requires at least three couplings. Among them, the trilinear coupling 
\begin{align}
\mathcal{L}_{\text{trilinear}}
= -\,VS_{R}{S_{L}}^\dagger H +\textrm{h.c.}~,
\end{align}
appears the most nontrivial. To accommodate it, we can extend the SM with a $SU(3)_L\times SU(2)_R$ global symmetry\footnote{The same symmetry has been studied in the CHM with top seesaw mechanism \cite{Cheng:2013qwa, Chung:2022avf} for a different purpose.}. Introducing a scalar $\Phi$ that transforms as a $(3,2)$ under the global symmetry, we find that the scalars $H$, $S_L$, $S_R$ together with a singlet $S_V$ can reside in the multiplet as
\begin{equation}
\Phi=(3,2)
\xrightarrow{\text{under SM}}
\begin{pmatrix}
1_0  & 1_{Q_F-\frac{2}{3}} \\ 
2_{\frac{1}{6}-Q_F}  & 2_{-\frac{1}{2}} \\ 
\end{pmatrix}
=
\begin{pmatrix}
S_V^*  & S_R^* \\ 
S_L  & H \\ 
\end{pmatrix},
\end{equation}
where the hypercharge was assigned to the desired value as shown in Eq. \eqref{hypercharge}. This assignment will be explained in the next section with a well-motivated origin.

The global symmetry allows for a $SU(3)_L\times SU(2)_R$ symmetric Mexican hat potential given by
\begin{align}
V(\Phi)
=-\mu^2 |\Phi^\dagger \Phi| + \lambda |\Phi^\dagger \Phi|^2~.
\end{align}
If only the singlet $S_V$ gets a nontrivial VEV with $\langle S_V\rangle=\sqrt{\mu^2/\lambda}$, we can get the desired trilinear coupling
\begin{align}
V(\Phi)\supset V \left(S_RS_L^\dagger H\right)+ \textrm{h.c.}~,
\end{align}
with the coefficient $V=2\lambda \langle S_V\rangle$. However, $S_L$ and $S_R$ will then be massless Nambu-Goldstone bosons. The complete potential should also include symmetry breaking terms, especially the SM gauge interactions. When such terms are included, two things happen: First, the loop-induced potential by the gauge interactions will preserve $\langle \Phi\rangle = \langle S_V\rangle=\sqrt{\mu^2/\lambda}$ as $S_V$ is a SM singlet. Second, both $S_L$ and $S_R$ will get a mass from the loop-induced potential.\footnote{If the charge $Q_F=2/3$, then $S_R$ will become chargeless and the argument fails. However, we can still recover it by assuming another $U(1)'$ gauge symmetry with the similar charge but $Q_F\neq 2/3$.}

In the fermion sector, we extend the SM content with a vector-like fermion $F$, according to the following $SU(3)_L\times SU(2)_R$ charges
\begin{equation}\label{fermions}
Q_L=
\begin{pmatrix}
F_L  \\ t_L \\ b_L \\
\end{pmatrix},
\quad
Q_R=
\begin{pmatrix}
F_R  \\ t_R \\
\end{pmatrix}~,
\end{equation}
where $Q_L$ is a triplet under $SU(3)_L$ and $Q_R$ is a doublet under $SU(2)_R$. We can then write down the Yukawa coupling between $Q$ and $\Phi$ as
\begin{align}
\mathcal{L}_{\text{Yukawa}}& =-y\, \bar{Q}_L \,\Phi\, Q_R \nonumber\\
&\supset  - y_{L}\bar{q}_L S_L F_R 
- y_{R}\bar{t}_R S_R F_L+\textrm{h.c.}~,
\end{align}
which includes the two Yukawa couplings we need with relation $y=y_L=y_R^*$. The Lagrangian also includes $ y\, \bar{F}_R S_V F_L$, which generates a mass for the vector-like fermion $F$. However, it also introduces an unwanted tree-level Yukawa coupling $ y\, \bar{q}_L H t_R$. The value of $y$ is too large and some modifications are required to get a realistic model.

\subsection{NJL model with an $SU(3)_L\times SU(2)_R$ global symmetry for the strong sector}

To get a realistic model, we separate the contents into two sectors, a strong sector and a weak sector, where the strong sector is responsible for the strong couplings and  the weak sector includes the SM matter contents. Starting with the strong sector, we first introduce a new set of fermions
\begin{equation}
Q'_L=
\begin{pmatrix}
F'_L  \\ t'_L \\ b'_L \\
\end{pmatrix},
\quad
Q'_R=
\begin{pmatrix}
F'_R  \\ t'_R \\
\end{pmatrix}~.
\end{equation}
To get the required strong couplings, a non-perturbative origin is expected. We consider a strong interaction among them mediated by massive gauge bosons with mass $M'$, which can arise from an asymptotically-free broken gauge symmetry. At scales below $M'$, we can integrate out the massive gauge bosons and get an effective four-fermion vertex term
\begin{align}
\mathcal{L}_{\text{eff}}&=-\frac{g'^2}{M'^2}
\left(\bar{Q'}_{L,i}\gamma^\mu T^a_{ij} {Q'}_{L,j}\right)
\left(\bar{Q'}_{R,i}\gamma_\mu T^a_{ij} {Q'}_{R,j}\right) \nonumber\\
&\supset\frac{g'^2}{M'^2}
\left({\bar{Q'}_L}^{~i}{Q'}_{R,i}\right)
\left({\bar{Q'}_R}^{~j}{Q'}_{L,j}\right),
\end{align}
where $g'$ is the coupling between the gauge bosons and $Q'$. If the coupling is strong enough, a fermion condensate will be formed, which can be described by a bound state given by
\begin{equation}
\bar{Q}'_RQ'_L=
\begin{pmatrix}
\bar{F'_R}\,F'_L & \bar{t'_R}\,F'_L \\
\bar{F'_R}\,q'_L & \bar{t'_R}\,q'_L\\
\end{pmatrix}
=
\begin{pmatrix}
S_V^*  & S_R^* \\ 
S_L  & S_H \\ 
\end{pmatrix},
\end{equation}
which carries the same gauge symmetries as the scalar field $\Phi$ from the last section. Indeed, the scalar field $\Phi$ turns out to be the bound state formed by $Q'_L$ and ${Q}'_R$, which is a natural origin for scalars in a strongly coupled theory.

Using the fermion bubble approximation, we can obtain the effective Lagrangian at a scale $\mu<M'$ by integrating out the fermion field components. The effective Lagrangian at the new scale $\mu$ will then be given by
\begin{align}
\mathcal{L}_\Phi=~&|\partial \Phi|^2-\tilde{M}(\mu)^2|\Phi|^2-\tilde{\lambda}(\mu)|\Phi|^4\nonumber\\
&-\tilde{y}(\mu) \,\bar{Q}'_L\,\Phi\, Q'_R+\textrm{h.c.}~,
\end{align}
where the coefficients are given by (defining $\ln(M'^2/\mu^2)=C$)
\begin{align}
&\tilde{M}(\mu)^2=\left(\frac{4\pi}{\sqrt{NC}}\frac{M'}{g'}\right)^{2}\left(1-\frac{g'^2}{g_c^2}+\frac{g'^2\,\mu^2}{\,g_c^2M'^2}\right),\nonumber\\
&\tilde{\lambda}(\mu)=\frac{16\pi^2}{NC}~,\quad
\tilde{y}(\mu)=\frac{4\pi}{\sqrt{NC}}~.
\end{align}
To generate the desired potential, we want it to be in the broken phase, i.e. $g'>g_c$. Together with the loop-induced potential from the SM gauge interaction (or some new $U(1)'$ gauge interaction), we get a VEV for $S_V$ as
\begin{equation}
\langle S_V\rangle=\langle \Phi\rangle \sim \sqrt{\frac{-\tilde{M}^2}{2\tilde{\lambda}}}\sim f'\sqrt{\frac{g'^2}{g_c^2}-1}~,
\end{equation}
where $f'\equiv M'/g'$ is the symmetry breaking scale of the strong dynamics. Now in the strong sector, we obtain
\begin{align}
\mathcal{L}_\Phi \supset&
~2\,\tilde{\lambda}\,\langle S_V\rangle ~(S_RS_L^\dagger S_H)\nonumber\\
&- \tilde{y}\,\bar{q}'_L S_L F'_R 
- \tilde{y}\,\bar{t}'_R S_R F'_L+\textrm{h.c.}~,
\end{align}
which is similar to the desired terms we need for the simplified scalar model. However now the terms are generated from a strongly coupled theory, where the couplings are naturally large. Also, the masses of $S_L$ and $S_R$ are generated from a loop-induced potential, but they can be much lighter in comparison to $M'$ due to the nature of pNGBs.

\subsection{Connecting the SM Higgs and top quarks with the strong sector}

Finally, we need to connect the strong sector with the SM content in the weak sector. Starting with the fermion sector, besides SM fermions, we still need a vector-like fermion $F$ with a mass $M_F$ as shown in \eqref{fermions}. We introduce two new extended $SU(2)$ gauge symmetries, $SU(2)_{L'}$ for the LH top quark and $SU(2)_{R'}$ for the RH top quark. The gauge symmetries are broken at the scales $f_L$ and $f_R$, respectively. Introducing two $SU(2)_{L'}$ doublets (the $SU(2)_{R'}$ case is analogous)
\begin{equation}
\psi_{q_L}=
\begin{pmatrix}
q'_L  \\ q_L \\
\end{pmatrix},
\quad
\psi_{F_R}=
\begin{pmatrix}
F_R  \\ F'_R \\
\end{pmatrix}\,,
\end{equation}
the $SU(2)_{L'}$ interaction will introduce an effective term as
\begin{align}
\mathcal{L}_{\text{eff}}&=-\frac{1}{f_L^2}
\left(\bar{\psi}_{q_L}\gamma^\mu T^a {\psi}_{q_L}\right)
\left(\bar{\psi}_{F_R}\gamma_\mu T^a {\psi}_{F_R}\right)\nonumber\\
&\supset\frac{1}{f_L^2}
\left({\bar{F}'_R}{q}'_{L}\right)\left({\bar{q}_L}{F}_{R}\right)
~\to~ y_{L}\bar{q}_L S_L F_R~,
\end{align}
where the desired Yukawa coupling is generated once the fermions in the strong sector form the bound states. The generic estimation for the Yukawa coupling is 
\begin{equation}
y_L\sim\frac{4\pi}{\sqrt{NC}}~f'^2 \times \frac{1}{f_L^2}
=\frac{4\pi}{\sqrt{NC}} \frac{f'^2}{f_L^2}
\end{equation}
where $f'$ is the generic VEV of the bound state. Switching to $SU(2)_{R'}$, we get
\begin{equation}
y_R\sim\frac{4\pi}{\sqrt{NC}}~f'^2 \times \frac{1}{f_R^2}
=\frac{4\pi}{\sqrt{NC}} \frac{f'^2}{f_R^2}
\end{equation}
If $f'\sim f_L \sim f_R$, then generically we get a large Yukawa coupling
\begin{equation}
y_L\sim y_R \sim \frac{4\pi}{\sqrt{NC}}~.
\end{equation}
Therefore, even though the top quarks and the vector-like fermion $F$ are not joining the strong interaction directly, we still get the desired large Yukawa couplings.

Concerning the Higgs sector, there is already a Higgs-like scalar field $S_H$ and all we need to introduce is a mixing term between it and SM Higgs $H^\dagger S_H$. Following its bound state nature $S_H=\bar{t}'_Rq'_L$, the mixing can be introduced by a coupling between the SM Higgs and its constituent fermion fields. As mentioned before, the whole mechanism is used to assist a model like SUSY or CHM, so we consider the two possibilities.

First, if the SM Higgs is elementary as in SUSY models, then the mixing can be generated through the Yukawa coupling between Higgs and the fermions in the strong sector as
\begin{align}
\mathcal{L}_{\text{}}=-y'\bar{q}'_L H t'_R ~\to~ y'f'^2 H^\dagger S_H
\end{align}
By integrating out the heavy $S_H$, we can then reproduce 
\begin{align}
\mathcal{L}_\text{trilinear}  =V (S_RS_L^\dagger H) +\textrm{h.c.}~,
\end{align}
with
\begin{align}
V\sim 2\,\tilde{\lambda}\,\langle S_V\rangle y' \frac{f'^2}{M_H^2}
\end{align}
where $M_H$ is the mass of $S_H$. The value of the trilinear coupling is thus controlled by the new Yukawa coupling $y'$ which can be small and plays the role of suppressing the top Yukawa coupling from a generic strong coupling down to $\mathcal{O}(1)$.

\begin{figure}[tbp]
\centering
\includegraphics[width=0.4\textwidth]{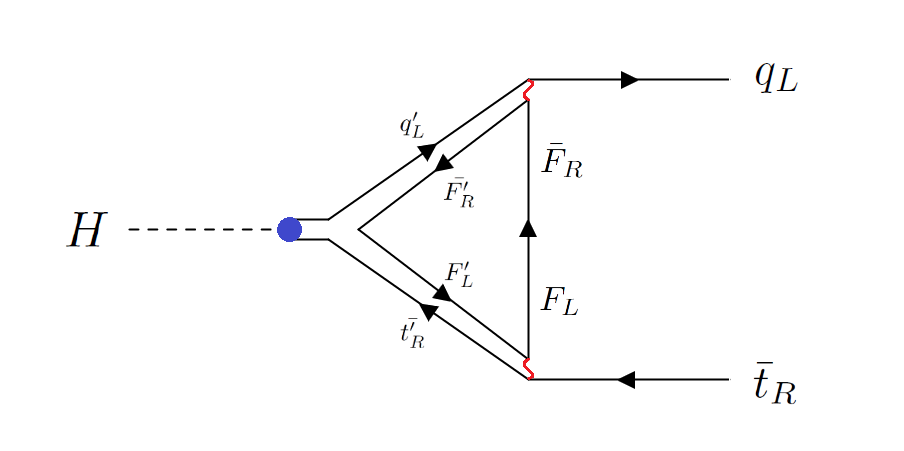}
\caption{Feynman diagram of the loop-generated top Yukawa coupling from a strongly coupled UV theory.\label{vertexUV}}
\end{figure}

In CHMs, the SM Higgs is itself composite, combined of $\psi_L$ and $\psi_R$. In order to mix it with the bound state $S_L$, a similar construction for the extended gauge symmetry is required as
\begin{align}
\mathcal{L}_{\text{eff}}&=-\frac{1}{f_{E}^2}
\left(\bar{\psi}_{L}\gamma^\mu T^a {q}'_{L}\right)
\left(\bar{\psi}_{R}\gamma_\mu T^a {t}'_{R}\right)\nonumber\\
&\supset\frac{1}{f_{E}^2}
\left({\bar{\psi}_R}{\psi}_{L}\right)\left({\bar{q}'_L}{t}'_{R}\right)
~\to~ \frac{f^2f'^2}{f_{E}^2}H^\dagger S_H~,
\end{align}
where $f_E$ is the scale of the extended gauge symmetry and $f$ is the breaking scale in the CHMs. Again by intergrating out the heavy $S_H$, we can get the trilinear coupling with coefficient
\begin{align}
V\sim 2\,\tilde{\lambda}\,\langle S_V\rangle \frac{f^2f'^2}{f_{E}^2M_H^2}~.
\end{align}
The overall structure of the top Yukawa vertex is shown in Fig. \ref{vertexUV}, where the scalars are now replaced by the bound states of fermions in the strong sector. The red line represents the gauge bosons of the extended gauge symmetry and the blue point is the mechanism to connect the Higgs boson to the strong sector.

\end{document}